%% file: main.tex
\documentclass[reprint, superscriptaddress, aps, prl, nofootinbib]{revtex4-2}

\usepackage{tabularx}
\usepackage{xcolor}
\usepackage{bm}
\usepackage{comment}
\usepackage{graphicx}
\usepackage{multirow}   
\usepackage{array}
\usepackage{booktabs} 
\usepackage[normalem]{ulem}
\usepackage{amsmath,amsthm}
\usepackage{mathtools}
\usepackage{amssymb}
\usepackage{acro}
\usepackage[utf8]{inputenc}
\usepackage[T1]{fontenc}
\usepackage{url}
\usepackage{verbatim}
\usepackage[pdfencoding=auto]{hyperref}
\hypersetup{colorlinks=true, citecolor=blue, urlcolor=blue, linkcolor=blue}

\input{acronyms}

\def\rr{{\bm r}}

\def\bAA{{\bm A}}
\def\bAAh{{\hat{\bm A}}}
\def\ll{{\bm l}}
\def\nn{{\bm n}}
\def\mm{{\bm m}}
\def\kk{{\bm k}}

\def\zz{{\bm z}}

\def\BB{{\bm B}}
\def\BBt{{\tilde{\bm B}}}
\def\BBh{{\hat{\bm B}}}

\def\ww{{\bm w}}
\def\vx{{\vec {x}}}
\def\vy{{\vec {y}}}
\def\vz{{\vec {z}}}

\newcommand{\vn}{\bm{n}}
\newcommand{\vl}{\bm{l}}
\newcommand{\vm}{\bm{m}}
\newcommand{\Stab}{\mathrm{Stab}}
\newcommand{\Fix}{\mathrm{Fix}}
\newcommand{\CG}{\mathcal{C}}
\newcommand{\calK}{\mathcal{K}}
\newcommand{\R}{\mathbb{R}}
\newcommand{\Sym}[1]{S_{#1}}
\newcommand{\tA}{\tilde{A}}
\newcommand{\tB}{\tilde{B}}
\newcommand{\hA}{\hat{A}}
\newcommand{\hB}{\hat{B}}
\newcommand{\OO}{\mathrm{O}(3)}

\newcommand{\Leq}[1]{Eq.~(L#1)}
\newcommand{\Leqs}[2]{Eqs.~(L#1)--(L#2)}
\DeclareMathOperator{\spn}{span}
\DeclareMathOperator{\tr}{tr}

\def\ptp{{    \begin{smallmatrix}
    \otimes^{lm}\vspace{-0.06cm}\\
    \odot_{k\hspace{0.155cm}}
    \end{smallmatrix} }}
    
\begin{document}

\title{Tensor-reduced atomic density representations}

\author{James P. Darby$^*$}
\affiliation{Warwick Centre for Predictive Modelling, School of Engineering, University of Warwick, Coventry, CV4 7AL, UK}
\affiliation{Engineering Laboratory, University of Cambridge, Cambridge, CB2 1PZ UK}

\author{D\'avid P. Kov\'acs$^*$}
\affiliation{Engineering Laboratory, University of Cambridge, Cambridge, CB2 1PZ UK}

\author{Ilyes Batatia}
\affiliation{Engineering Laboratory, University of Cambridge, Cambridge, CB2 1PZ UK} \affiliation{ENS Paris-Saclay, Universit\'e Paris-Saclay, 91190 Gif-sur-Yvette, France}

\author{Miguel~A.~Caro}
\affiliation{Department of Electrical Engineering and
Automation, Aalto University, FIN-02150 Espoo, Finland}

\author{Gus L. W. Hart}
\affiliation{Department of Physics and Astronomy, Brigham Young University, Provo, Utah, 84602, USA}

\author{Christoph Ortner}
\affiliation{Department of Mathematics, University of British Columbia, 1984 Mathematics Road, Vancouver, BC, Canada V6T 1Z2}

\author{G\'abor Cs\'anyi}
\affiliation{Engineering Laboratory, University of Cambridge, Cambridge, CB2 1PZ UK}

\date{\today}

\begin{abstract}

Density based representations of atomic environments that are invariant under Euclidean symmetries have become a widely used tool in the machine learning of interatomic potentials, broader data-driven atomistic modelling and the visualisation and analysis of materials datasets. The standard mechanism used to incorporate chemical element information is to create separate densities for each element and form tensor products between them. This leads to a steep scaling in the size of the representation as the number of elements increases.
Graph neural networks, which do not explicitly use density representations, escape this scaling by mapping the chemical element information into a fixed dimensional space in a learnable way.
By exploiting symmetry, we recast this approach as tensor factorisation of the standard neighbour density based descriptors and, using a new notation, identify connections to existing compression algorithms. 
In doing so, we form compact tensor-reduced representation of the local atomic environment whose size does not depend on the number of chemical elements, is systematically convergable and therefore remains applicable to a wide range of data analysis and regression tasks.

{\bf Erratum: } {\it The proof of completeness of the tensor factorisation we give in the Supplement is only valid up to correlation order three. We explain this in detail in an appended Erratum. A more detailed investigation of the consequences of this observation is given in a forthcoming work.}
\end{abstract}

\maketitle
\def\thefootnote{*}\footnotetext{These authors contributed equally.}

Over the past decade, machine learning methods for studying atomistic systems have become widely adopted~\cite{deringer2021origins, Duat22022DielsAlder, kapil2022first}. Most of these methods utilise representations of local atomic environments that are invariant under relevant symmetries; typically rotations, reflections, translations and permutations of equivalent atoms~\cite{musil2021physics, zhang2018end}. Enforcing these symmetries allows for greater data efficiency during model training and ensures that predictions are made in a physically consistent manner. There are many different ways of constructing such representations which are broadly split into two categories: (i) descriptors based on internal coordinates, such as the Behler-Parrinello Atom-Centered Symmetry Functions~\cite{Behler2007ACSF}, and (ii) density-based descriptors such as \ac{SOAP}~\cite{bartok2013representing} or the bispectrum~\cite{bartok2010gaussian,THOMPSON2015SNAP}, which employ a symmetrised expansion of $\nu$-correlations of the atomic neighbourhood density ($\nu=2$ for SOAP and $\nu=3$ for the bispectrum). A major drawback of all these representations is that their size increases dramatically with the number of chemical elements $S$ in the system. For instance, the number of features in the linearly complete \ac{ACE}~\cite{ACE_ralf, DUSSON2022} descriptor which unifies, extends and generalises the aforementioned representations, scales as $S^\nu$ for terms with correlation order $\nu$ (i.e. a body order of $\nu+1)$. This poor scaling severely restricts the use of these representations in many applications. For example, in the case of machine learned interatomic potentials for systems with many (e.g. more than 5) different chemical elements, the large size of the models results in memory limitations being reached during parameter estimation as well as significantly reducing evaluation speed.

Multiple strategies to tackle this scaling problem have been proposed including element weighting~\cite{gastegger2018wacsf, artrith2017ACSF_weight} or  embedding the elements into a fixed small dimensional space~\cite{willatt2018feature, gubaev2019accelerating}, directly reducing the element-sensitive correlation order~\cite{darby2022compressing}, low-rank tensor-train approximations for lattice models~\cite{kostiuchenko2019impact} and data-driven approaches for selecting the most relevant subset or combination of the original features for a given dataset~\cite{NICE, goscinski2021optimal, zeni2021compact}. 
A rather different class of machine learning methods are \acp{MPNN}~\cite{gilmer2017neural, schnet}. Instead of constructing full tensor products, these models also embed chemical element information in a fixed size latent space using a learnable transformation $\mathbb{R}^S \rightarrow \mathbb{R}^K$ where $K$ is the dimension of the latent space, and thus avoid the poor scaling with the number of chemical elements. Recently these methods have achieved very high accuracy~\cite{nequip, batatia2022design, MACE2022}, strongly suggesting that the true complexity of the relevant chemical element space does not grow as $S^\nu$.

In this paper we introduce a general approach for significantly reducing the scaling of density-based representations like \ac{SOAP} and \ac{ACE}. We show that by exploiting the tensor structures of the descriptors and applying low-rank approximations we can derive new tensor-reduced descriptors which are systematically convergeable to the original full descriptor limit. We also verify this with numerical experiments on real data. We also show that there is a natural generalisation to compress not only the chemical element information but also the radial degrees of freedom, yielding an even more compact representation. When fitting interatomic potentials for organic molecules and high entropy alloys, we achieve a ten-fold reduction in the number of features required when using linear (\ac{ACE}) and nonlinear kernel models (\ac{SOAP}-GAP). We also fit a linear model to a dataset with 37 chemical elements which would be infeasible without the tensor-reduced features.
%

All many-body density based descriptors can be understood in terms of \ac{ACE} \cite{ACE_ralf}. In ACE, the first step in describing the local neighbourhood $\mathcal{N}(i) = \{j : r_{ij} < r_{\rm cut}\}$ around atom $i$ is forming the one-particle basis $	\phi_{znlm}(\rr_{ij}, Z_j)$ as a product of radial basis functions $R_n$, spherical harmonics $ Y_{l}^m$ and an additional element index shown in Eq.~\eqref{eq:ACE_one_p_basis},  where $\rr_{ij}$ and $Z_j$ denote the relative position and atomic number of neighbour $j$. Permutation invariance is introduced by summing over neighbour atoms in Eq.~\eqref{eq:ACE_atomic_basis} after which $(\nu+1)$-body features are formed in Eq.~\eqref{eq:ACE_product_basis} by taking tensor products of the atomic basis $A_{i,znlm}$ with itself $\nu$ times. Finally,  Eq.~\eqref{eq:ACE_basis} shows how the product basis $\bAA_{i, \zz\nn\ll\mm}$ is rotationally symmetrised using the generalised Clebsch-Gordon coefficients $\mathcal{C}^{\ll\eta}_\mm$, where $\eta$ enumerates all possible symmetric couplings for a given $\ll$ resulting in invariant features~\cite{ACE_ralf, DUSSON2022, NICE}.
\begin{align}
	\phi_{znlm}(\rr_{ij}, Z_j) &= R_n(r_{ij}) Y_{l}^m(\hat{\rr}_{ij}) \delta_{zZ_j},\label{eq:ACE_one_p_basis}\\
	A_{i,znlm} &= \sum_{j \in \mathcal{N}(i)} \phi_{znlm}(\rr_{ij}, Z_j),   \label{eq:ACE_atomic_basis}\\ 
    \bAA_{i, \zz\nn\ll\mm} &= \prod_{t = 1}^\nu A_{i, z_t n_t l_t m_t}     \label{eq:ACE_product_basis} \\
    \BB_{i, \zz\nn\ll \eta} & = \sum_{\mm} \mathcal{C}^{ \ll \eta}_{\mm} \bAA_{i,\zz\nn\ll\mm} \label{eq:ACE_basis}
\end{align}
A linear ACE model can the be fit to an invariant atomic property $\varphi_i$ as 
\begin{equation} \label{eq:standardace} 
    \varphi_i
    = 
    \sum_{\zz \nn \ll \eta}
    c_{\zz \nn \ll \eta} \BB_{i, \zz \nn \ll \eta}
\end{equation}
where $c_{\zz \nn \ll \eta}$ are the model parameters and for practical reasons the expansion is truncated using $\nu_\mathrm{max}$, $l_\mathrm{max}$ and $n_\mathrm{max} := N$, the number of radial basis functions. Note that as $\BB_{i, \zz\nn\ll\eta}$ is invariant under ${(z_a, n_a, l_a) \xleftrightarrow{} (z_b, n_b, l_b)}$ symmetrically equivalent terms are usually omitted from Eq.~\eqref{eq:standardace}, again see~\cite{ACE_ralf, DUSSON2022, NICE} for the details. For some $\ll$ there is no invariant component of $\bAA_{i, \zz\nn\ll\mm}$ so that $\mathcal{C}^{ \ll \eta}_{\mm}=0$ and the corresponding trivial $\BB_{i, \zz\nn\ll \eta}$ are removed from Eq.~\eqref{eq:standardace}.

Crucially, the tensor product in Eq.~\eqref{eq:ACE_product_basis} causes the number of features (and therefore the number of model parameters) to grow rapidly as $\mathcal{O}(N^\nu S^\nu)$. Previous work \cite{willatt2018feature, goscinski2021optimal, ACE_ralf} has reduced this to $\mathcal{O}(K^\nu)$ by first embedding the chemical and radial information into $K$ channels (Eq.~\eqref{eq:embedding}), then taking a full tensor product across the $\bar{A}_{i, klm}$ (Eq.~\eqref{eq:k_tensor_product}).
\begin{align}
    \bar{A}_{i, klm} &= \sum_{zn} W^{k}_{zn} A_{i, z n l m}, \quad k=1\ldots K \label{eq:embedding}\\
    \bar{\bAA}_{i, \kk\ll\mm} &= \prod_{t=1}^\nu \bar{A}_{i, k_t l_t m_t} \label{eq:k_tensor_product}
\end{align}
This approach is also used in Moment Tensor Potentials \cite{gubaev2019accelerating, novikov2020mlip} and in Gaussian Moment descriptors~\cite{zaverkin2020gaussian}. The embedding can be identified in Eq. 3 of ref. \cite{novikov2020mlip}, where $\mu$ indexes the embedded channels and $\nu$ is similar to $l$ in ACE. Then taking tensor products across the embedded channels corresponds to forming products of the moments $M_{\mu \nu}$. In general, the embedding weights are optimised either before or during fitting \cite{willatt2018feature, novikov2020mlip, ACE_ralf} with the latter causing the models to be non-linear. 

We propose a principled approach to further reduce the size of the basis to $\mathcal{O}(K)$ which can be understood from two different angles. First, we identify the model parameters $\bm{c}_\eta \equiv c_{\zz \nn \ll \eta}$ in Eq.~\eqref{eq:standardace} as a symmetric tensor, invariant under $(z_a, n_a, l_a) \xleftrightarrow{} (z_b, n_b, l_b)$, which can be expanded as a  sum of products of rank-1 tensors~\cite{comon2008symmetric} as,
\begin{equation}
	\bm{c}_\eta = \sum_{k=1}^K \lambda_{k\eta}  \underset{\nu \text{ times}}{\underbrace{\ww_k \otimes \ww_k \dots  \otimes \ww_k}},
	\label{eq:SOP}
\end{equation}
or in component form
\begin{equation}
	c_{\zz \nn \ll \eta} = \sum_k^{K} \lambda_{k\eta} \prod_{t=1}^\nu W^{k}_{z_t n_t l_t}
	\label{eqn:sym_expansion_main}
\end{equation}
where $W^{k}_{z_t n_t l_t}$ are the components of $\ww_k$. This expansion is exact for finite $K$, as $\bm{c}$ is finite due to basis truncation, and is equivalent to eigenvalue decomposition of a symmetric matrix when $\nu=2$.  Note that we choose to use the same weights $W^{k}_{z_t n_t l_t}$ for all $\nu$ and $\eta$, which significantly reduces the number of weights that need be specified. In practice, we can choose to expand over the $\zz\nn$ or $\zz$ indices only, see SI for details, and then substitute the expansion into Eq.~\eqref{eq:standardace} as
\begin{align}
    \varphi_i &\approx \sum_{k\ll\eta} \lambda_{k\ll\eta} \left[\sum_{\mm} \mathcal{C}^{\ll \eta}_{\mm} \sum_{\zz \nn} \prod_{t=1}^\nu W^{k l_t}_{z_t n_t} A_{i,z_t n_t l_t m_t}\right]\label{eq:decoupled_ace_mid}\\
    &=\sum_{k\ll\eta} \lambda_{k\ll\eta} \left[\sum_{\mm} \mathcal{C}^{\ll \eta}_{\mm} \prod_{t=1}^\nu \tilde{A}_{i, k l_t m_t}  \right]\\
    &=\sum_{k\ll\eta} \lambda_{k\ll\eta} \BBt_{i,k\ll\eta} \label{eq:decoupled_ace_main}
\end{align}
where $\BBt_{i,k\ll\eta}$ are the new tensor reduced features and the approximation arises because in practice we truncate the tensor decomposition early. The key novelty is that only element-wise products are taken across the $k$ index of the embedded channels $\tilde{A}_{i, k l_t m_t}$ when forming the many-body basis, rather than a full tensor product, i.e. $k$ does not have a $t$ subscript in Eq.~\ref{eq:decoupled_ace_mid} (see Table~\ref{tab:summary} for the full definitions). For completeness, we note that applying this tensor reduction to the elements only and using $K=2$ is equivalent to the element-weighting strategies used in \cite{gastegger2018wacsf, artrith2017ACSF_weight, uhrin2021through}.

There are multiple natural strategies for specifying the embedding weights $W^{kl}_{zn}$, including approximating a pre-computed $c_{\zz \nn \ll \eta}$ or treating the weights as model parameters to be estimated during the training process, as is done in MACE~\cite{MACE2022}. Here we investigate using random weights as a simpler alternative. This ensures that Eq.~\ref{eq:decoupled_ace_main} remains a linear model and allows the $\BBt_{i,k\ll\eta}$ to be used directly in other tasks such as data visualisation. 
\begin{table*}[t!]
    \caption{The density projection $\vec{A}_i$ are viewed as vectors with a composite index $(z, n, l, m)$ whereas the embedded density projections $\bar{A}_i = (\bm{W}\vec{A}_i)$ etc. are indexed by $(k, l, m)$. The most general tensor reduced many body density projection (``Tensor sketch'') and its special case (``Tensor decomposition'') are shown together with their scaling with the number of radial basis functions, $N$, number of chemical elements, $S$ and number of embedding channels, $K$. The symbol $\ptp$ means full tensor product across $l$ and $m$ but element-wise product across $k$ whereas $\otimes$ indicates a full tensor product across all indices.}
    \label{tab:summary}
    \centering
    \begin{tabular}{cccc}
    \toprule
        Name & Product Basis & Index Notation  & Basis size \\
    \midrule
     ACE    & \hspace{0.5cm} $\bAA_i =  \vec{A}_i \otimes  \vec{A}_i \dots  \otimes  \vec{A}_i$ \hspace{0.5cm} &   $\bAA_{i, \zz\nn\ll\mm} = \prod_{t=1}^\nu A_{i, z_t n_t l_t m_t}$ & $\mathcal{O}(NS)^\nu$ \\
     \cite{ACE_ralf} & & &\\
    Embedding    & \hspace{0.5cm} $\bar{\bAA}_i =  (\bm{W}\vec{A}_i) \otimes  (\bm{W}\vec{A}_i) \dots  \otimes  (\bm{W}\vec{A}_i)$ \hspace{0.5cm} &   $\bar{\bAA}_{i, \kk\ll\mm} = \prod_{t=1}^\nu \bar{A}_{i, k_t l_t m_t}$ & $\mathcal{O}(K^\nu)$ \\
    \cite{willatt2018feature, goscinski2021optimal, gubaev2019accelerating, ACE_ralf, bochkarev2022efficient}& & $\bar{A}_{i, k_t l_t m_t} = \sum_{zn} W^{k_t}_{zn} A_{i, z n l_t m_t}$ & \\
    \\
    Tensor decomposition   & \hspace{0.5cm} $\tilde{\bAA}_i =  (\bm{W}\vec{A}_i) \ptp  (\bm{W}\vec{A}_i) \dots  \ptp (\bm{W}\vec{A}_i)$ \hspace{0.5cm} &   $\tilde{\bAA}_{i, k\ll\mm} = \prod_{t=1}^\nu \tilde{A}_{i, k l_t m_t} $   & $\mathcal{O}(K)$ \\
    \cite{MACE2022} & & $\tilde{A}_{i, k l_t m_t} = \sum_{zn} W^{kl_t}_{zn} A_{i, z n l_t m_t}$ & \\
    \\
    Tensor sketch   & \hspace{0.5cm} $\hat{\bAA}_i =  (\bm{W}^1\vec{A}_i) \ptp  (\bm{W}^2\vec{A}_i) \dots  \ptp  (\bm{W}^\nu\vec{A}_i)$ \hspace{0.5cm} &   $\hat{\bAA}_{i, k\ll\mm} = \prod_{t=1}^\nu \hat{A}_{i t, k l_t m_t} $ &   $\mathcal{O}(K)$ \\
    & & $\hat{A}_{it,  k l_t m_t} = \sum_{zn} W^{t k}_{zn} A_{i, z n l_t m_t}$  & \\
    \bottomrule
    \end{tabular}
\end{table*}

 We now show that the resulting tensor-reduced features can also be understood from the perspective of directly compressing the original $\BB_{i, \zz\nn\ll\eta}$ features. \ac{RP}~\cite{dasgupta2013experiments, bingham2001random} is an established technique where general high dimensional feature vectors $\{\vx_1, \dots, \vx_N\}\subset \mathbb{R}^{d}$ are compressed as $\tilde{x}_i = \bm{W}\vx_i \in \mathbb{R}^{K}$, with the entries of the matrix $\bm{W}$ being normally distributed. This approach is simple, offers a tuneable level of compression and is underpinned by the Johnson-Lindenstrauss Lemma~\cite{johnson1984extensions} which bounds the fractional error made in approximating $\vx^T_i \vx_j$ by $ \tilde{x}^T_i \tilde{x}_j$. \ac{RP} can also be used to reduce the cost of linear models, with a closely related approach recently used in \cite{browning2022gpu}. In \ac{CLSR}~\cite{kaban2013new, ahmed2017big, maillard2009compressed} features are replaced by their projections, thus reducing the number of model parameters. Loosely speaking, the approximation errors incurred in \ac{CLSR} (and \ac{RP} in general) are expected to decay as $1/\sqrt{K}$ and we refer to refs. \cite{kaban2013new, maillard2009compressed, hsu2012random} for more details. The drawback of \ac{RP} is that it requires the full feature vector to be constructed so that applying \ac{RP} to ACE would not avoid the unfavourable $\mathcal{O}(N^\nu S^\nu)$ scaling. We propose using tensor sketching~\cite{liu2022tensor, ahle2020oblivious, pham2013fast, woodruff2014sketching}, as used in refs. \cite{holzmuller2022framework, zaverkin2022exploring} to approximate the Neural Tangent Kernel, instead of \ac{RP}. For vectors with tensor structure $\vx = \vy \otimes \vz$ where $\vx \in \mathbb{R}^{d_1d_2}$, $\vy \in \mathbb{R}^{d_1}$ and $\vz \in \mathbb{R}^{d_2}$, the Random Projection $\bm{W}\vx$ can be efficiently computed directly from $\vy$ and $\vz$ as
\begin{equation}
\bm{W} \vx = (\bm{W'}\vy \,) \odot (\bm{W''} \vz \,).
\label{eq:TS}
\end{equation}
where $\odot$ denotes the element-wise (Hadamard) product. Similarly, the ACE product basis can be tensor sketched, across the $zn$ indices,  as
\begin{equation}
    \hat{\bAA}_i =  (\bm{W}^1\vec{A}_i) \ptp  (\bm{W}^2\vec{A}_i) \dots  \ptp  (\bm{W}^\nu\vec{A}_i)
\end{equation}
where $\ptp$ denotes taking the tensor product over the upper indices $lm$ and the element-wise product over the lower index $k$ and $\bm{W}^1, \bm{W}^2$ etc. are i.i.d. random matrices, see SI for details. The $\hat{\bAA}_{i, k\ll\mm}$ can then be symmetrised as in Eq.~\ref{eq:ACE_basis} yielding
\begin{equation}
    \hat{\BB}_{i, k\ll\eta} = \sum_m \mathcal{C}^{\ll\eta}_\mm \prod_{t=1}^\nu \hat{A}_{i t, k l_t m_t} 
    \label{eq:decoupled_TS}
\end{equation}
where $\hat{A}_{i t, k l_t m_t}$ is defined more precisely in Table~\ref{tab:summary}. Finally, we note that because the embedded channels are independent, the error in approximating inner products using the average across $K$ channels is expected to decrease as $1/\sqrt{K}$, just as with standard \ac{RP}. Based on this, we conjecture that similar bounds derived for the errors made in \ac{CLSR} may also apply here. A summary comparing standard ACE, element-embedding and tensor-reduced ACE is given in Table~\ref{tab:summary}, where it is clear that the features derived using tensor decomposition are equivalent to the tensor-sketched features with the choice of using equal weights in each factor.

We now turn to numerical results and first demonstrate that the tensor-reduced features are able to efficiently and completely describe a many-element training set. We consider a dataset comprised of all symmetry inequivalent fcc structures made up of 5 elements with up to 6 atoms per unit cell~\cite{hart2008algorithm}. A set of features is complete on this dataset if the design matrix for a linear model fit to total energies has full (numerical) row rank, where each row corresponds to a different training configuration. 

\begin{figure}[h!]
    \centering
    \includegraphics[width=0.5\textwidth]{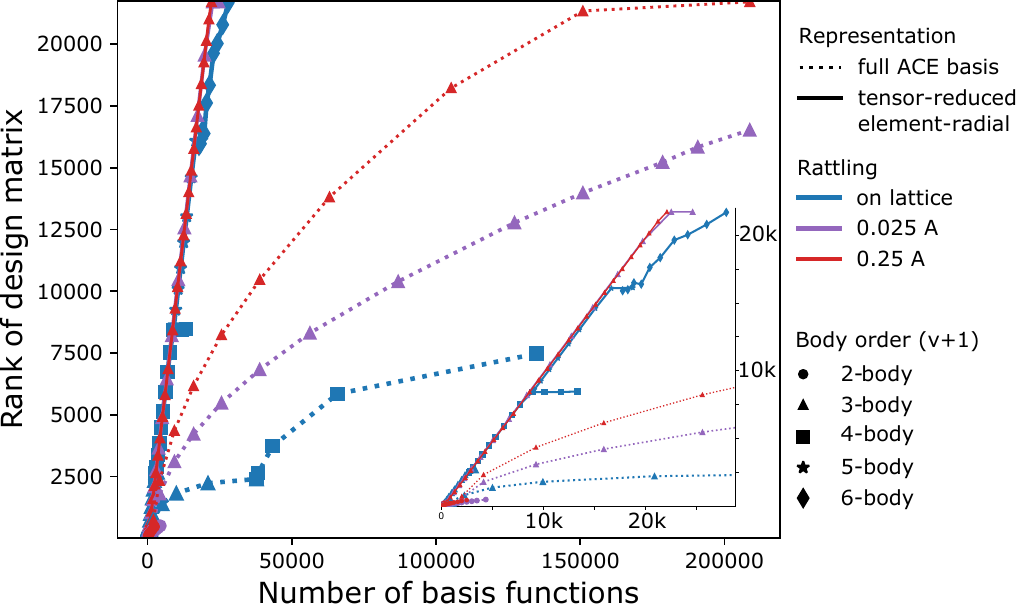}
    \caption{The row rank of the design matrix as a function of basis set on a dataset of all symmetry inequivalent fcc lattices of 5 chemical elements and unit cell sizes of up to 6 atoms. The inset zooms in on the $x=y$ region. The colours denote the size of the rattling around the crystal sites of an ideal fcc lattice, the markers correspond to increasing body-order of the basis and the line-style compares the fully-coupled and tensor-reduced ACE versions.  
    \label{fig:rank_test}
    }
\end{figure}

Figure~\ref{fig:rank_test} shows the numerical rank of the design matrix as a function of the basis set. At a given correlation order the standard ACE basis set is grown by increasing the polynomial degree, and the tensor-reduced basis set is enlarged by increasing $K$, the number of independent channels. In both cases once the rank stops increasing at the given correlation order we increment $\nu$.
The colors in Fig.~\ref{fig:rank_test} correspond to three different geometrical variations: blue contains on-lattice configurations only whilst in magenta and red the atomic positions have been perturbed by a random Gaussian displacement with mean 0 and standard deviation of 0.025 and 0.25 \AA, respectively. The dotted lines corresponds to the standard ACE basis, whereas the solid lines corresponds to the tensor-reduced version from Eq.~\eqref{eq:decoupled_ace_main}. 
Although the standard ACE basis can always achieve full row rank since it is a complete linear basis, it does this very inefficiently. In contrast, the row rank using the tensor-reduced basis grows almost linearly. Thus the tensor-reduced basis, having removed unnecessary redundancies, still retains the expressive power of the full basis.

\begin{figure}[hb!]
    \centering
    \includegraphics[width=0.4\textwidth, trim={0 0 0 1.45cm}, clip]{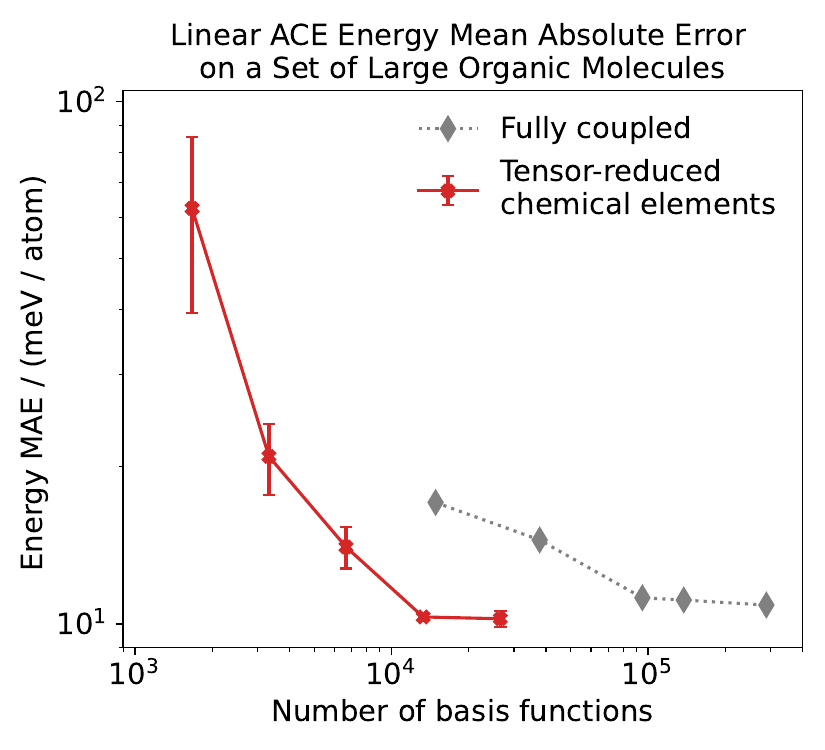}
    \caption{Convergence of the energy errors of organic molecules on the independent test set with respect to the number of basis functions is shown for linear models using standard and  tensor-reduced (tensor decomposition) \ac{ACE} features. Error bars show the standard error in the mean, computed across 5 fits using independently chosen random weights.
    \label{fig:ACE_QMugs}
    }
\end{figure}

Next, we fit a linear \ac{ACE}~\cite{kovacs2021} model on a training set of 400 different organic molecules of size 19-168 atoms, randomly selected from the QMugs dataset~\cite{isert2022qmugs} that are made up of 10 different chemical elements (H, C, N, O, F, P, S, Cl, Br, I). The conformers were created by running GFN2-xTB~\cite{bannwarth2019gfn2}  800~K NVT molecular dynamics for 1~ps starting from a published minimum energy structure. The test set is composed of 1,000 \emph{different} molecules sampled the same way. This is a small-data regime task ($\sim$60,000 scalar targets) that is particularly challenging due to the chemical and conformational diversity.  Figure~\ref{fig:ACE_QMugs} shows the convergence of the energy error with the number of basis functions for the fully coupled and the tensor-reduced \ac{ACE} models, both using $\nu_\mathrm{max}=3$ (4-body). By increasing the number of uncoupled channels $K$ we can converge the accuracy to the previous level, whilst reducing the size of the model by a factor of 10. 

\begin{figure}[htbp!]
    \centering
    \includegraphics[width=0.4\textwidth]{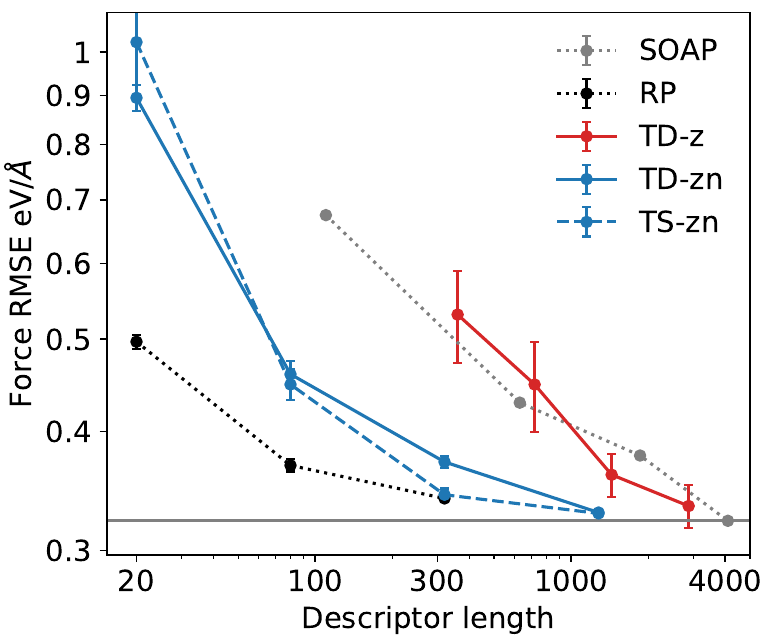}
    \caption{Convergence of force RMSE on the high-entropy alloy test set as a function of descriptor length for various GAP models.   Error bars show the standard error in the mean, computed across 5 fits using independently chosen random weights. The tensor reduction is denoted using TD for tensor-decomposition~\eqref{eq:decoupled_ace_main}, TS for tensor-sketching~\eqref{eq:decoupled_TS} and the reduction is performed over the indices after the hyphen. }
    \label{fig:HEA_fit}
\end{figure}

The tensor reduction techniques proposed in this work can be directly applied to other density based descriptors and used with alternative models. To demonstrate this we
used the tensor-reduced SOAP power spectrum to fit \ac{GAP} \cite{bartok2010gaussian} models to the quinary high-entropy alloy dataset~($\sim$400,000 scalar targets) from ref. \cite{byggmastar2021modeling}. The  descriptors were computed using turbo-soap~\cite{caro2019optimizing} and fitting was performed using the GAP code~\cite{gapfit}. To provide a baseline for comparison, reference models were fit using the full power spectrum evaluated with a varying number of basis functions; $n_{\text{max}}=2l_{\text{max}}$. All compressed descriptors were constructed using the largest values of $n_\mathrm{max}=8, l_\mathrm{max}=4$, see Supplemental Material~\cite{SI} for details. The force errors achieved on the independent test set are shown as a function of descriptor length in Fig.~\ref{fig:HEA_fit}. For this dataset, mixing only the chemical elements did not help, which is likely due to the minimal savings made as a result of the low correlation order $\nu$. However, using the fully element-radial tensor-reduced descriptor allowed for similar accuracy to be achieved using approximately 10 times fewer features. Furthermore, the errors achieved with all compressed descriptors are shown to converge to the error achieved with the full \ac{SOAP} vector they were derived from. Finally, in the high accuracy regime, descriptor length > 300,  models using the tensor-sketch element-radial reduction matched the accuracy achieved using full \ac{RP}. 

The information content of these SOAP based descriptors was also measured using the information imbalance \cite{glielmo2021ranking}. It was found that the fully element-radial tensor reduced descriptors offered comparable data efficiency to a random projection and that taking an element-wise product across embedded channels outperformed taking a full tensor product, see Supplemental Material~\cite{SI} for details.

Finally, we assess the effect of optimising the embedding weights using the MACE architecture~\cite{MACE2022}. A typical MACE model is a 2-layer message passing network which utilises tensor-reduced ACE features to efficiently represent equivariant body-ordered messages. Normally, the embedding weights are optimised using backpropagation along with all other model parameters, but it is possible to fix these weights to random values. As such, whilst a multi-layer MACE is a complex non-linear network model, a single-layer MACE with frozen embedding weights is equivalent to a linear model using tensor-reduced ACE features~\cite{batatia2022design}. Furthermore, within MACE the embedding weights used in the tensor decomposition, see Table \ref{tab:summary}, are approximated as $W^{kl_t}_{zn}$ = $U_{zn}V^{kl_t}_n$ so that the element weights $U_{zn}$ and the radial weights $V^{kl_t}_n$ can be optimised or frozen independently. 
 
We use the HME21 dataset~\cite{Takamoto2022HME21, takamoto2022towards} ($\sim$1,700,000 scalar targets) for this test due to its exceptional diversity, both chemically with 37 elements, and structurally with configurations including isolated molecules, bulk crystals, surfaces, clusters and disordered materials. We fit several models (see Supplemental Material~\cite{SI} for details), and the energy and force errors on the independent test set are summarised in Table \ref{tab:acac}. Strikingly, using random element embedding weights leads to almost no degradation in model accuracy compared to using optimised element embedding weights. In contrast, randomizing both the element and radial embedding weights lead to significantly larger prediction errors.
A two-layer model with optimised weights achieves state of the art accuracy on HME21, showing the power of the message passing architecture and further highlighting the effectiveness of the tensor-reduced features. 

\begin{table}[h!]
    \caption{ \textbf{Mean absolute errors on the HME21 dataset.}
    Energy ($E$, meV) and force ($F$, eV/\AA) errors of models. The labels, (1) or (2), of the MACE models indicate the number of message passing layers. MACE errors are an average over three seeds.}
    \vspace{5pt}
    \label{tab:acac}
    \centering
    \resizebox{0.48\textwidth}{!}
     {%
    \begin{tabular}{lm{0.5cm}ccccccc}
    \toprule
        &  & MACE $(1)$\, &   MACE $(1)$ \,& MACE $(1)$ & \, MACE $(2)$ \, & \, TeaNet~\cite{Takamoto_2022} \, & \, NequIP~\cite{nequip}  \\
        &$U_{zn}$   & opt  & rand & rand  & opt &   &         &    \\
        &$V^{kl_t}_{n}$  & opt  & opt & rand & opt &  & &\\
        \midrule
        \multirow{2}{*}{}
                      & $E$ & 53.8   & 54.7  & 254.8 & \textbf{16.5} & 19.6 &    47.8  \\
                      & $F$ & 0.194  &  0.193  & 0.451 & \textbf{140.2}  &  0.174 &  0.199    \\    
        \bottomrule
    \end{tabular}
    }
\end{table}

In conclusion, we introduced a tensor-reduced form of the \ac{ACE} basis for modelling symmetric functions of local atomic neighbour environments that eliminates the $\mathcal{O}\left(N^\nu S^\nu\right)$ scaling of the basis set size with the number of chemical elements and radial basis functions. Intuitively, the construction can be thought of as mixing the element and radial channels and then only coupling these channels to themselves when constructing the higher order many-body basis. We derived this new embedded basis from a symmetric tensor decomposition and explored its connection to tensor-sketching. We showed that this reduced basis is also systematic, and 
that in practice it can enable a 10-fold reduction in basis set size for diverse datasets with many elements, including organic molecules and high entropy alloys. When using in a 2-layer message passing network, MACE~\cite{MACE2022}, the tensor-decomposition yields state of the art performance.

\begin{acknowledgements}
DPK acknowledges support from AstraZeneca and the EPSRC. GC acknowledges discussion with Boris Kozinsky.  JPD and GC acknowledge support from the NOMAD Centre of Excellence, funded by the European Commission under grant agreement 951786. We used computational resources of the UK HPC service ARCHER2 via the UKCP consortium and funded by EPSRC grant EP/P022596/1. CO acknowledges support of the Natural Sciences and Engineering Research Council [Discovery Grant IDGR019381] and the New Frontiers in Research Fund [Exploration Grant GR022937].
\end{acknowledgements}

\bibliographystyle{PRL}
\bibliography{references}
\clearpage

\clearpage
\widetext
\section*{Supplemental Material}

\subsection{Symmetric Tensor Decomposition of \texorpdfstring{$c_{\zz\nn\ll\eta}$ }{parameters}}
We start by relabelling the parameter tensor $c_{\zz\nn\ll\eta}$  of Eq.~\eqref{eq:standardace} to $c_{\nn\ll\eta}$ where the element index has been incorporated into $\nn = ((z_1, n_1), (z_2, n_2), ... (z_\nu, n_\nu))$. Then we remove the lexicographical ordering of $(\zz, \nn, \ll)$ and retain \textbf{all} terms in the tensor-product so that there are redundancies in $\BB_{i, \nn\ll\eta}$. The advantage of this is that the generalised Clebsch-Gordon coefficients do not depend on $\nn$, which simplifies this analysis. Next we note that $c_{\nn\ll\eta}$  is symmetric under $(n_i, l_i) \xleftrightarrow{} (n_j, l_j)$, but not $n_i \xleftrightarrow{} n_j$ or $l_i \xleftrightarrow{} l_j$. We can now use the fact that a symmetric tensor can be expanded in terms of rank-1 tensors, i.e.
\begin{equation}
	\bm{c} = \sum_k \lambda_k \ww^k \otimes \ww^k \otimes \dots  \otimes \ww^k
\end{equation}
or, rewritten in component form,  
\begin{equation}
	c_{\nn \ll \eta}  \sim \sum_k^{K} \lambda_{k\eta} [\ww^k]^{l_1}_{n_1} \cdots  [\ww^k]^{l_\nu}_{n_\nu} 
	\label{eqn:sym_expansion}
\end{equation}
where the components of $\ww^k$ are indexed with the tuple $(n_i, l_i)$ as $[\ww^k]^{l_i}_{n_i} = w^{k l_i}_{n_i}$.  Note, that the same set of $\ww^k$ are used to expand all $c_{\nn \ll \eta}$, rather than having a separate set for each $(\ll, \eta)$ tuple. Consequently the expansion for each $(\ll, \eta)$ will converge slower, although we stress that it is still systematic, but crucially we require far fewer embedded one particle basis functions. We then insert this expression into 
\begin{equation} 
	\varphi
	= 
	\sum_{\ll \mm \eta} \mathcal{C}^{ \ll \eta}_{\mm} \sum_{\nn} c_{\nn\ll \eta} \bAA_{\nn\ll\mm}.
	\label{eqn:21}
\end{equation}
and get 
	\begin{align} \label{eq:160}
	\sum_{\nn} c_{\nn\ll \eta} \bAA_{\nn\ll\mm}
	&= 
	\sum_{\nn} \sum_k \lambda_{k\eta}
	\prod_{t = 1}^\nu w^{k l_t}_{n_t} A_{n_t l_t m_t} \\ 
	&= 
	\sum_k \lambda_{k\eta}
	\sum_{\nn} 
	\prod_{t = 1}^\nu w^{k l_t}_{n_t} A_{n_t l_t m_t} \\ 
	&= 
	\sum_k \lambda_{k\eta}
	\prod_{t = 1}^\nu \sum_n w^{k l_t}_{n} A_{n l_t m_t} \\ 
	&=
	\sum_k \lambda_{k\eta}
	\prod_{t = 1}^\nu \tilde{A}_{k, l_t, m_t} 
\end{align}
where
\begin{align}
	\tilde{A}_{k, l_t, m_t} 
	&= 
	\sum_n w^{k l_t}_{n} A_{n l_t k_t} \\ 
	&= 
	\sum_n  w^{k l_t}_{n}
	\sum_j R_{n l_t}(r_{j}) Y_{l_t}^{m_t}(\hat\rr_{j}) \\ 
	&= 
	\sum_j  \bigg[ \sum_n w^{k l_t}_{n}R_{n l_t}(r_{j}) \bigg] 
	Y_{l_t}^{m_t}(\hat\rr_{j}) \\ 
	&=: 
	\sum_j  \tilde{R}_{k l_t}(r_{j})
	Y_{l_t}^{m_t}(\hat\rr_{j}).    
	\label{eqn:decoupled_AA}
\end{align}
and  $ \tilde{R}_{k l_t}(r_{j}) = \sum_n w^{k l_t}_{n}R_{n l_t}(r_{j})$ is a mixed element-radial basis function. Subbing Eq.~\eqref{eqn:decoupled_AA} back into Eq.~\eqref{eqn:21} we arrive at
\begin{align}
	\varphi &= \sum_{k\eta} \lambda_{k\eta}  \BBt_{k\eta} \label{eqn:JLT}\\ 
	\BBt_{k\eta} &= \sum_{\ll \mm} \mathcal{C}^{ \ll \eta}_{\mm} \tilde{\bAA}_{k \ll \mm} \label{eqn:Bq}, \\ 
	\tilde{\bAA}_{k \ll \mm} &= \prod_t \tilde{A}_{k l_t m_t}.
\end{align}
Where each basis function $\BBt_{k\eta}$ involves a sum over all $\ll$ tuples at the given correlation order.  An alternative option is to decompose $c_{\nn\ll \eta}$ approximately as 

\begin{equation}
	c_{\nn \ll \eta} \sim \sum_k^{k_\mathrm{max}} \lambda_{k\ll\eta} [\ww^{kl_1}]_{n_1} \cdots  [\ww^{kl_\nu}]_{n_\nu},
	\label{eqn:l_expansion}
\end{equation}
where the difference is that $[\ww^k]^{l_i}_{n_i} \rightarrow [\ww^{kl_i}]_{n_i}$.  Proceeding as above results in the identical analysis as before, except that now the basis functions become
\begin{align}
	\varphi &= \sum_{k\ll} \lambda_{k\ll\eta}  \BBt_{k\ll\eta}\\ 
	\BBt_{k\ll\eta} &= \sum_{\mm} \mathcal{C}^{ \ll \eta}_{\mm} \tilde{\bAA}_{k \ll \mm} \label{eqn:Bql}
\end{align}
and $\tilde{\bAA}_{k \ll \mm}$ is as above. Both are valid expansions and comparing Eq.~\eqref{eqn:Bq} and Eq.~\eqref{eqn:Bql} we see that $\BBt_{k\eta} = \sum_\ll \BBt_{k\ll\eta}$. In this work we choose to investigate using $\BBt_{k\ll\eta}$ as every computed $\BBt_{k\ll\eta}$ is used separately, so that the model is more flexible for a similar evaluation cost. We note that using $\BBt_{k\eta}$ might result in a greater level of model accuracy for a given basis set size but leave this investigation to future work.
\subsection{Uncoupled Basis as a Tensor Sketch}
Here we investigate whether inner products are preserved when moving from the full \ac{ACE} basis to the uncoupled basis. Precisely, we investiage whether $\sum_{\ll\eta} \BBt_{i,k\ll\eta}  \BBt_{j, k\ll\eta} $  from Eq.~\eqref{eq:decoupled_ace_main} and $\sum_{\ll\eta} \BBh_{i,k\ll\eta} \BBh_{j, k\ll\eta} $ from Eq.~\eqref{eq:decoupled_TS} are unbiased estimators for $\sum_{\zz\nn\ll\eta} \BB_{i,\zz\nn\ll\eta}  \BB_{j, \zz\nn\ll\eta}$ from Eq.~\eqref{eq:standardace} when the weights $W^{kl}_{zn}$ are drawn from a symmetric distribution with mean 0. We start by combining the element and radial indices so that $\BB_{\zz\nn\ll\eta} \rightarrow \BB_{\nn\ll\eta}$.  First we note that

\begin{align}
	\sum_{\nn\ll\eta} \BB_{i, \nn\ll\eta}  \BB_{j, \nn\ll\eta}   &= \sum_{\nn\ll\eta} \left( \sum_\mm \mathcal{C}^{ \ll \eta}_{\mm} \bAA_{i, \nn\ll\mm} \right)
	\left( \sum_{\mm'} \mathcal{C}^{ \ll \eta}_{\mm'} \bAA_{j, \nn\ll\mm'} \right) \\
	& = \sum_{\ll\mm\mm'\eta} \mathcal{C}^{ \ll \eta}_{\mm}\mathcal{C}^{ \ll \eta}_{\mm'} \left( \sum_\nn \bAA_{i, \nn\ll\mm}  \bAA_{j, \nn\ll\mm'}  \right) \label{eqn:37}
\end{align}
 and similarly
\begin{align}
	\sum_{\kk\ll\eta} \BBh_{i, \kk\ll\eta} \BBh_{j, \kk\ll\eta}  &= \sum_{\kk\ll\eta} \left( \sum_\mm \mathcal{C}^{ \ll \eta}_{\mm} \bAAh_{i, \kk\ll\mm} \right)
	\left( \sum_{\mm'} \mathcal{C}^{ \ll \eta}_{\mm'} \bAAh_{j, \kk\ll\mm'} \right) \\
	& = \sum_{\ll\mm\mm'\eta} \mathcal{C}^{ \ll \eta}_{\mm}\mathcal{C}^{ \ll \eta}_{\mm'} \left( \sum_\kk \bAAh_{i, \kk\ll\mm}  \bAAh_{j, \kk\ll\mm'}  \right) \label{eqn:39}
\end{align}
where the sum over $\kk$ is over randomly chosen tuples $(k_1, k_2, ... k_\nu)$ where $k_i \neq k_j$. Comparing  Eq.~\eqref{eqn:37} and Eq.~\eqref{eqn:39} we see that it is sufficient to check that 
\begin{equation}
	\sum_\kk  \mathbb{E}[\bAAh_{i, \kk\ll\mm}  \bAAh_{j, \kk\ll\mm'}] =  \sum_\nn \bAA_{i, \nn\ll\mm}  \bAA_{j, \nn\ll\mm'}
\end{equation}
Next we note that every $\kk$ tuple is independent, so we need only check one term in the sum on the left. Inserting the definition of $\bAAh_{\kk\ll\mm}$ we find that
\begin{align}
		\bAAh_{i, \kk\ll\mm}  \bAAh_{j, \kk\ll\mm'} & = \prod_t^\nu \tilde{A}^{k_t}_{i, l_t m_t}  \tilde{A}^{k_t}_{j, l_t m'_t} \\
		& = \prod_t^\nu \left(\sum_n W^{k_tl_t}_{ n} A_{i, n l_t m_t} \right) \left( \sum_{n'} W^{k_tl_t}_{ n'} A_{j, n' l_t m'_t} \right) \\
		& = \sum_{\nn \nn'} \prod_t^\nu W^{k_tl_t}_{n'_t} W^{k_tl_t}_{n_t} A_{i, n_t l_t m_t} A_{j, n'_t l_t m'_t} \\
		& = \sum_{\nn \nn'}  \prod_t^\nu  A_{i, n_t l_t m_t} A_{j, n'_t l_t m'_t}  \prod_t^\nu W^{k_tl_t}_{n_t} W^{k_tl_t}_{n'_t}\\
\end{align}
Next we take the expectation value
\begin{align}	
	\mathbb{E}\left[\bAAh_{i, \kk\ll\mm}  \bAAh_{j, \kk\ll\mm'}\right] & = \sum_{\nn \nn'}  \prod_t^\nu  A_{i, n_t l_t m_t} A_{j, n'_t l_t m'_t} \mathbb{E}\left[ \prod_t^\nu W^{k_tl_t}_{n_t} W^{k_tl_t}_{n'_t} \right]\\ \label{eqn:W_product}
	&=  \sum_{\nn \nn'}   \prod_t^\nu  A_{i, n_t l_t m_t} A_{j, n'_t l_t m'_t} \mathbb{E}\left[W^2\right]^\nu \delta_{\nn \nn'} \\
	&= \mathbb{E}\left[W^2\right]^\nu \sum_\nn \prod_t^\nu A_{i, n_t l_t m_t} A_{j, n_t l_t m'_t} \\
	&= \mathbb{E}\left[W^2\right]^\nu  \sum_\nn \bAA_{i, \nn\ll\mm} \bAA_{j, \nn\ll\mm}  \label{eqn:ACE_inner_prod}
\end{align}
where $\mathbb{E}\left[ W^{k_i l_i}_{n_i} W^{k_j l_j}_{n'_i} \right] = \delta_{k_i k_j} \delta_{l_i l_j} \delta_{n_i n'_i} \mathbb{E}\left[W^2\right]$ because the $W^{k_i l_i}_{n_i}$ are i.i.d. random variables with mean 0. Eq.~\eqref{eqn:ACE_inner_prod} is the result we require, up to a constant factor of $ \mathbb{E}\left[W^2\right]^\nu$ which can be absorbed into the definition of $\BBt_{\kk\ll\eta}$. 
\newline\newline
The analysis for $\BBt_{k\ll\eta}$  follows in the same way except that we have $\mathbb{E}\left[ \prod_i^\nu W^{k_il_i}_{n_i} W^{k_il_i}_{n'_i} \right] \rightarrow \mathbb{E}\left[ \prod_i^\nu W^{kl_i}_{n_i} W^{kl_i}_{n'_i}\right]$ in Eq.~\eqref{eqn:W_product}. If all $l_i$ in $\ll$ are unique then  $\mathbb{E}\left[ \prod_i^\nu W^{kl_i}_{n_i} W^{kl_i}_{n'_i}\right] = \prod_i^\nu \mathbb{E}\left[W^2\right] \delta_{n_i n'_i}$. However if any of the $l_i$ are repeated then there are additional terms which adjust the expectation value e.g. $\ll = (2, 3, 2)$,
\begin{align}
	\mathbb{E}\left[ W^{k,2}_{n_1} W^{k,2}_{n'_1}  \cdot W^{k,3}_{n_2} W^{k,3}_{n'_2} \cdot W^{k,2}_{n_3} W^{k,2}_{n'_3}\right] =
	\\ \mathbb{E}[W^2]^3 \left( \delta_{n_1n_1'} \delta_{n_2n_2'} \delta_{n_3n_3'}  + \delta_{n_1 n_3 } \delta_{n_2 n'_2} \delta_{n'_1 n'_3}  + \delta_{n_1 n'_3 } \delta_{n_2 n'_2} \delta_{n'_1 n_3}    \right) \\ + \left(\mathbb{E}[W^2] \mathbb{E}[W^4] - 3\mathbb{E}[W^2]^3 \right) \delta_{n_2 n'_2} \delta_{n_1, n'_1}  \delta_{n_1, n_3}  \delta_{n_1, n'_3} 
\end{align}
This means that $\BBt_{k\ll\eta}$ from Eq.~\eqref{eq:decoupled_ace_main} is not a true tensor sketch.

\clearpage
\subsection{Details of constructing the ACE basis}
\subsubsection{Rank test}
\paragraph{Numerical Rank}
The design matrix considered was constructed by evaluating all basis functions on all sites in the unit cell and summing up their values. This way we obtained a matrix $\Psi$ of the size $ (\# \text{basis} \times \# \text{configurations})$. The rank of $\Psi$ was computed by counting the number of non-zero singular values of the matrix with a tolerance of $n*\epsilon$, where $n$ is the smallest dimension of the matrix and $\epsilon$ is the machine epsilon of the type \texttt{Float64} in the Julia programming language $(2.220446049250313e-16)$.  

\paragraph{Details of the standard ACE basis}
For the ACE basis we used a cutoff radius of 10. The maximum degree of each body-order was increased gradually until the numerical rank stopped increasing. The other parameters of the basis are identical to the one used in \cite{kovacs2021} as implemented in \texttt{ACE.jl}

\paragraph{Details of the Tensor-reduced ACE basis}

For the Tensor-reduced ACE basis we also used a cutoff of 10 \AA. At each of the body order we used a maximum radial polynomial degree of 21, meaning that the radial functions were a random combination of the 21 radial polynomials. For the angular part we used maximum L of the spherical harmonics for each of the correlation orders as follows: $1 => 0, \quad 2 => 12,\quad 3 => 10,\quad 4 => 8,\quad 5 => 6$. The motivation for having lower L for higher correlation order stems from the expectation that higher body order contribution will be smoother. 

\subsubsection{Linear ACE models}

The models had a cutoff of 4.5 \AA. The models were fitted using linear least squares regression with Laplacian preconditioning regularization as detailed in Ref~\cite{kovacs2021}. The basis size was increased by gradually increasing the maximum polynomial degree of the ACE basis, with the inclusion of up to 4-body basis functions. The Tensor-reduced ACE basis also included up to 4-body features and was grown by increasing the number of embedding channels. 

\begin{figure}[h!]
    \centering
    \includegraphics[width=0.4\textwidth]{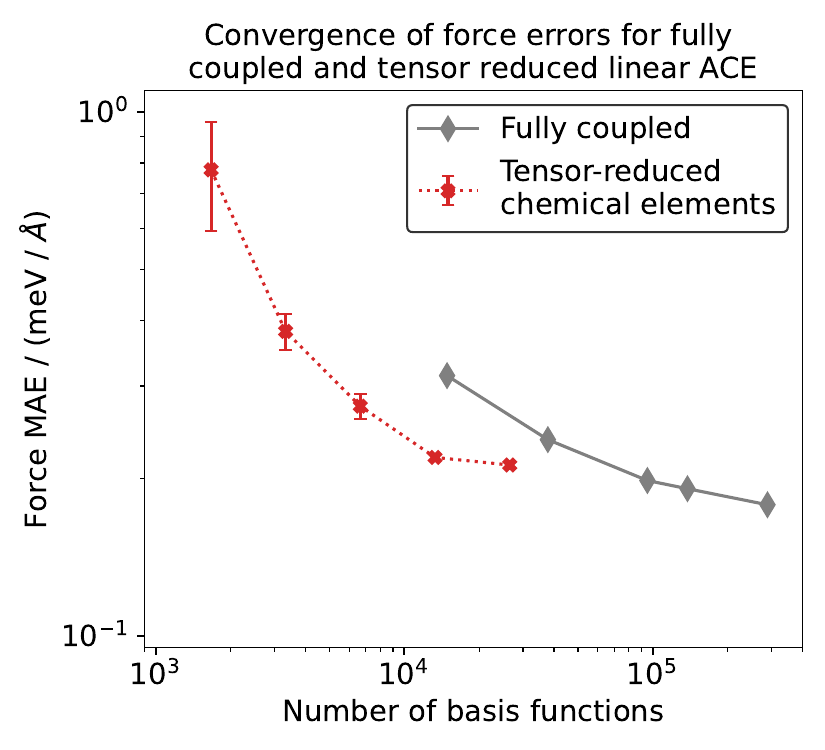}
    \label{fig:lin_ACE_Forces}
    \caption{Convergence of linear Tensor-reduced ACE force error to the fully coupled with increasing number of channels. Error bars show the standard error in the mean, computed across 5 fits using independently chosen random weights
    }
\end{figure}

We have also trained models by applying the tensor reduction to both the chemical element and the radial channels. In this case we find that the errors do not converge as rapidly resulting in higher errors with similar basis set size compared to the fully coupled basis set. For models with ca. 20,000 basis functions applying tensor reduction to chemical elements only gives 0.2 eV/\AA, chemical element - radial tensor reduction gives 0.5 eV/\AA, whereas the fully coupled basis set has an error of 0.3 eV/\AA.  This observation is common to the linear ACE and one-layer MACE (which is an alternative implementation of linear ACE), but is different from the behaviour of the kernel regression based radial-element reduced SOAP kernel model. We hypothesise that whilst the chemical element degrees of freedom can efficiently be decoupled the radial decoupling only works in the case of non-linear models. 

\clearpage

 \subsection{SOAP Information Imbalance}
Next, we investigate tensor-reduced forms of the SOAP power spectrum. The information imbalance,  introduced in ref. \cite{glielmo2021ranking}, provides a quantitative way to measure the relative information content of descriptors by comparing the pairwise distances between atomic environments in a given dataset. For a given environment, all other environments in the dataset are ranked (denoted by $r$) according to their distance from it using two different descriptors $A$ and $B$. The information imbalance is then defined as 
\begin{equation}
    \Delta_{B\rightarrow A} = \frac{2}{n_\mathrm{env}} \langle r_B| r_A=1 \rangle
\end{equation}
where $r_B| r_A=1$ is the rank according to $B$ of the nearest neighbour according to $A$ and the average $\langle r_B| r_A=1 \rangle$ is across all $n_\mathrm{env}$ environments in the dataset.  Defined as such, for large $n_\mathrm{env}$, $\Delta_{B\rightarrow A} \approx 0$ when $B$ contains the same information as $A$ and $\Delta_{B\rightarrow A} \approx 1$ if $B$ contains no information about $A$. By using nearest neighbour information, $\Delta_{B\rightarrow A}$ is sensitive only to the local neighbourhood around a given data point and is insensitive to any scaling of the distances, making it well suited to study non-linear relationships between different distance measures.

\begin{figure}[h!]
    \centering
    \includegraphics[width=0.45\textwidth]{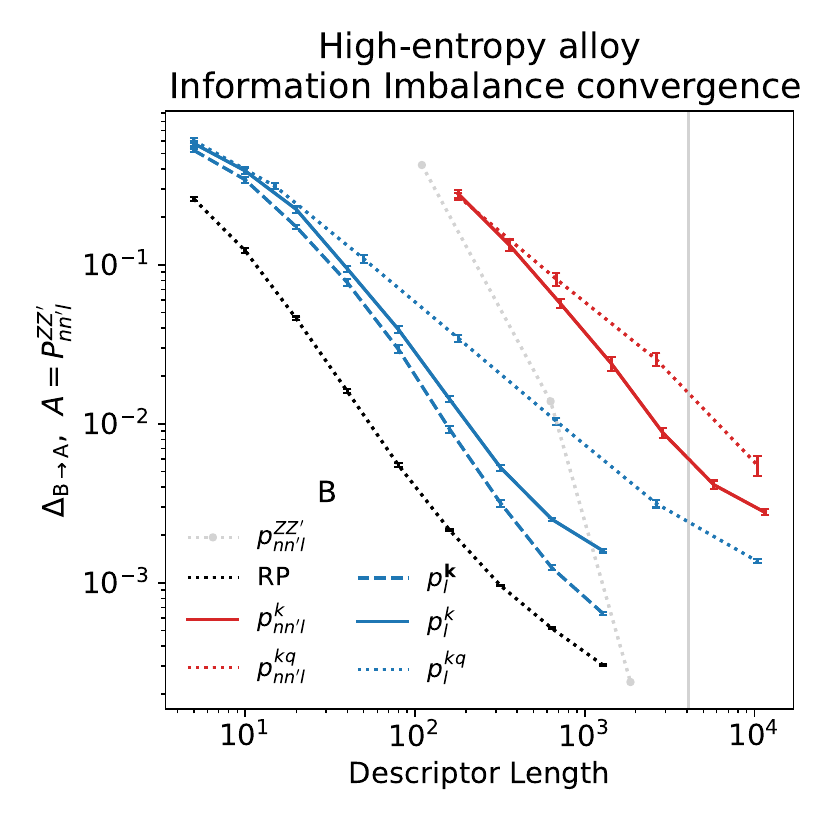}
    \caption{Convergence of the information imbalance, relative to the full power spectrum, for 1000 randomly chosen environments from liquid configurations in a high-entropy alloy dataset \cite{byggmastar2021modeling}, S=5. Element and element-radial mixing are indicated using red and blue respectively. Fully coupled channels are indicated with a dotted line whilst uncoupled channels are indicated with a solid (or dashed) line. The black dotted line corresponds to a random projection of the full power spectrum whilst the grey dotted line corresponds to $p^{zz'}_{nn'l}$ computed using fewer basis functions (while keeping the $n_\mathrm{max}=2l_\mathrm{max}$ relation). Error bars show the standard error in the mean across 10 randomly chosen sets of weights and the vertical grey line indicates the length of the reference power spectrum, $n_\mathrm{max}=8, l_\mathrm{max}=4$.}
    \label{fig:II_new}
\end{figure}

In applying the tensor reduction formalism to SOAP, we switch to its customary notation, i.e. $\mathbf{c}^z_{nl}$ is the neighbour density expansion (playing the role of the atomic basis $A$ of ACE) and $p^{zz'}_{nn'l} = \mathbf{c}^z_{nl}\cdot \mathbf{c}^{z'}_{n'l}$ for the power spectrum (corresponding to the product basis $\bm A$ in ACE for $\nu=2$, where the Clebsch-Gordan coefficients are all 1)\cite{DeringerCsanyi2021ChemRev}.
In Fig.~\ref{fig:II_new} the information imbalance between the full SOAP power spectrum and various tensor-reduced forms is shown as a function of descriptor length for a dataset of high-entropy alloy liquid environments containing 5 elements. New mixed element channels are constructed by randomly mixing element densities and denoted as ${\tilde{\mathbf{c}}^k_{nl} =\sum_z w^k_z \mathbf{c}^z_{nl}}$ whilst mixed element-radial channels are denoted as ${\hat{\mathbf{c}}^k_{l}=\sum_{z n} w^k_{z n} \mathbf{c}^z_{nl}}$.

The uncoupled power spectrum of mixed element and radial channels is then $p^{k}_l =\hat{\mathbf{c}}^k_{l} \cdot \hat{\mathbf{c}}^k_{l}$ (solid blue), as in  Eq.~\eqref{eq:decoupled_ace_main}. If only element channels are mixed, we obtain $p^{k}_{nn'l} =\tilde{\mathbf{c}}^k_{nl} \cdot \tilde{\mathbf{c}}^k_{n'l}$ (solid red). In analogy with Eq.~\eqref{eq:decoupled_TS}, using $p^{\mathbf{k}}_l = \hat{\mathbf{c}}^{k_1}_{l} \cdot \hat{\mathbf{c}}^{k_2}_{l}$ (dashed blue) marginally outperforms $p^k_l$ on this test, which is expected as $p^{\mathbf{k}}_l$ provides an unbiased estimator for the reference distances. In the same plot we show the performance of features obtained by constructing the full tensor product of the new mixed channels,  $p^{kq}_l =\hat{\mathbf{c}}^k_{l} \cdot \hat{\mathbf{c}}^q_{l}$ (mixing element and radial channels, dotted blue), and $p^{kq}_{nn'l} =\tilde{\mathbf{c}}^k_{nl} \cdot \tilde{\mathbf{c}}^q_{n'l}$ (mixing element channels only, dotted red). Both for the uncoupled and the full tensor product cases, mixing elements and radials together (blue lines) outperforms mixing elements only (red lines). 

A random projection~\cite{johnson1984extensions, bingham2001random, dasgupta2013experiments}  of $p^{zz'}_{nn'l}$ (black dashed) and the full power spectrum $p^{zz'}_{nn'l}$ itself but computed using fewer basis functions (grey dotted), which is a naive way of obtaining a shorter descriptor, are also included as references. Using the fully tensor-reduced power spectrum $p^{k}_l$ provides a performance only slighty inferior to random projection whilst avoiding constructing the full $p^{zz'}_{nn'l}$ as an intermediate step. Notably, using element-only mixing (red lines) is less efficient than creating a shorter descriptor simply by lowering the spatial resolution.  Equivalent results for the revised MD17~\cite{christensen2020role} dataset and a single element phosphorous dataset~\cite{deringer2020general} are shown below.

The \ac{SOAP} descriptors were computed using a radial cut-off of 5 \AA, $n_{\text{max}}=8$, $l_{\text{max}}=4$, $\sigma_{\text{at}}=0.5$ \AA and the weight of the central atom set to 0. Shorter versions of the full power spectrum were computed using $n_{\text{max}}=2l_{\text{max}}$ for $l_{\text{max}}=1, 2, 3$.

\begin{figure}[h!]
    \centering
    \includegraphics[width=0.42\textwidth]{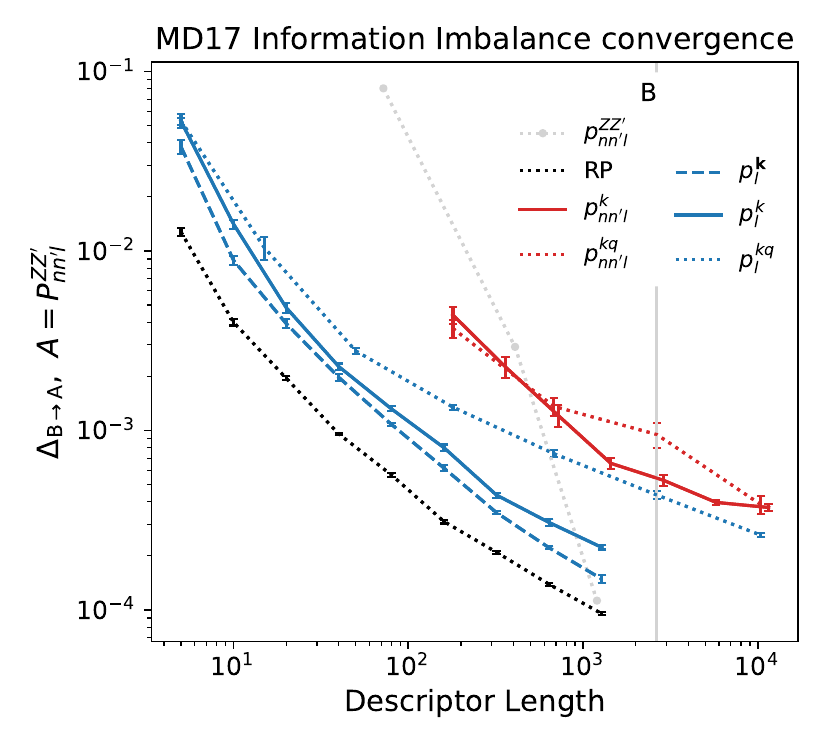}
    \label{fig:md17_II}
    \caption{Convergence of the information imbalance with respect to descriptor length for 2000 carbon environments randomly selected from the revised MD17 dataset~\cite{christensen2020role}. Element and element-radial mixing are indicated using red and blue respectively. Fully coupled channels are indicated with a dotted line whilst uncoupled channels are indicated with a solid (or dashed) line. The black dotted line is \ac{RP} on the full power spectrum. Error bars show the standard error in the mean computed across 10 (20 for the red lines) randomly chosen sets of weights. The grey  line corresponds to using the full power spectrum truncated using $n_\mathrm{max}=2l_\mathrm{max}$ for $l_\mathrm{max} = 1, 2, 3$ where $l_\mathrm{max}$=4 is the reference used for all comparisons.}
    \end{figure}

\begin{figure}[h!]
    \centering
    \includegraphics[width=0.42\textwidth]{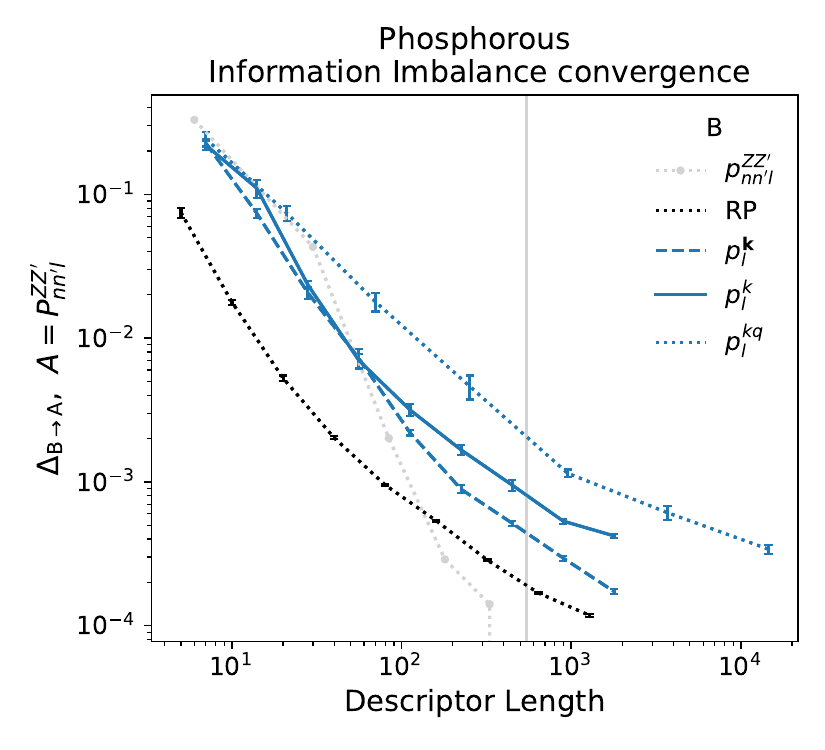}
    \caption{Convergence of the information imbalance, relative to the full \ac{SOAP} power spectrum, is shown as a function of descriptor length for a dataset of P environments. The dataset was constructed by extracting 2000 environments from the training dataset of ref. \cite{deringer2020general} using CUR decomposition and the same \ac{SOAP} parameters as the original work \cite{DeringerCsanyi2021ChemRev} were used. Fully coupled channels are indicated with a dotted line whilst uncoupled channels are indicated with a solid (or dashed) line. The black dotted line corresponds to a random projection of the full power spectrum. Error bars show the standard error in the mean computed across 10 randomly chosen sets of weights. The grey  line corresponds to using the full power spectrum truncated using $n_\mathrm{max}=2l_\mathrm{max}$ for $l_\mathrm{max} = 1, 2, 3, 4, 5$ where $l_\mathrm{max}=6$ is the reference used for all comparisons. }
    \end{figure}
    
\clearpage
\subsection{SOAP-GAP energy convergence}
The \texttt{gap\_fit} command used to fit one of the models is shown below. The same parameters were used for all models with only the \texttt{compress\_file} being changed. Note that for brevity only one example for each descriptor type has been included. For the actual models a two-body term was included for every pair of elements and a SOAP term was included for every element.

\small
\begin{verbatim}
at_file=db_HEA_reduced.xyz core_param_file=pairpot.xml core_ip_args={IP Glue} sparse_jitter=1e-8
    gp_file=ER_diag_normal_K=64_2.xml rnd_seed=999 default_sigma={ 0.002 0.1 0.5 0.0} 
gap={  {distance_2b cutoff=5.0 cutoff_transition_width=1.0 covariance_type=ard_se delta=10.0
    theta_uniform=1.0 n_sparse=20 Z1=42 Z2=42 add_species=F}  :    
{soap_turbo species_Z={23 41 42 73 74} l_max=4 n_species=5 rcut_hard=5 rcut_soft=4 basis=poly3gauss
    scaling_mode=polynomial radial_enhancement=0 covariance_type=dot_product add_species=F delta=1.0
    n_sparse=2000 sparse_method=cur_points zeta=6 compress_file=ER_diag_KK_normal-8-4-5-64_2.dat
    central_index=1 alpha_max={8 8 8 8 8} atom_sigma_r={0.5 0.5 0.5 0.5 0.5} 
    atom_sigma_t={0.5 0.5 0.5 0.5 0.5} atom_sigma_r_scaling={0.0 0.0 0.0 0.0 0.0} 
    atom_sigma_t_scaling={0.0 0.0 0.0 0.0 0.0} amplitude_scaling={0.0 0.0 0.0 0.0 0.0} 
    central_weight={1.0 1.0 1.0 1.0 1.0}}
\end{verbatim}

\begin{figure}[h!]
\centering
\includegraphics[width=0.45\textwidth]{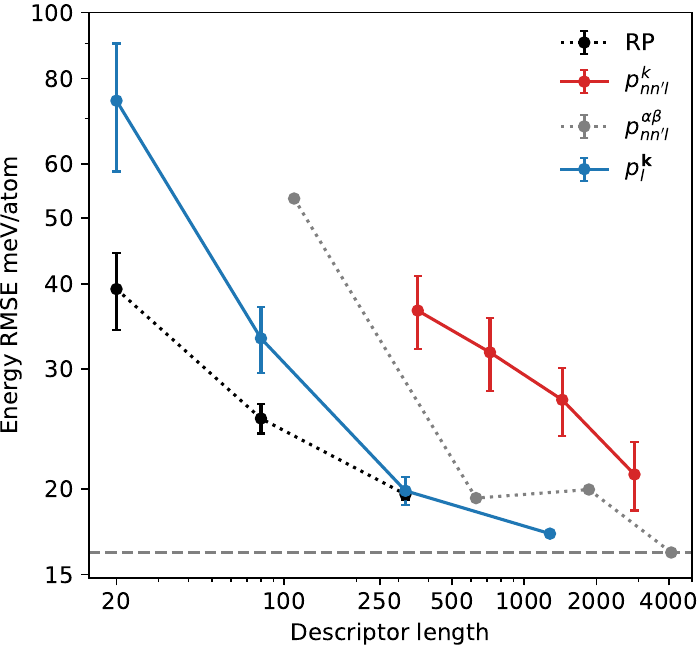}
\caption{Convergence of energy errors on the test set as a function of descriptor length for various GAP models. The grey line is full \ac{SOAP} using $n_\mathrm{max}=2l_\mathrm{max}$, for $l_\mathrm{max}=1, 2, 3, 4$. The black dotted line is \ac{RP} on the full power spectrum whilst the red and blue lines use a tensor-reduced features with element and element-radial mixing respectively. Error bars show the standard error in the mean, computed across 5 fits using independently chosen random weights.}
\end{figure}

\subsection{Details of the MACE models}

The MACE models were trained on NVIDIA A100 GPU in single GPU training. The training, validation and test split of the HME21 dataset \cite{Takamoto2022HME21, takamoto2022towards} were used. The data set was reshuffled after each epoch. We constructed three seperate models, two with one layer and one with two layers. On all models, we used 256 uncoupled feature channels, $l_{max}=3$ and pass invariant messages $L_{max}=2$ only. For the model with random weights, we froze the weights of the initial chemical embedding. For all models, radial features are generated using 8 Bessel basis functions and a polynomial envelope for the cutoff with p = 5. The radial features are fed to an MLP of size [64, 64, 64, 1024], using SiLU nonlinearities on the outputs of the hidden layers. The readout function of the first layer is implemented as a simple linear transformation. For the model with the second layer, the readout function of the second layer is a single-layer MLP with 16 hidden dimensions. We used a cutoff of 6 \AA . The standard weighted energy forces loss was used, with a weight of 1 on energies and a weight of 10 on forces \cite{MACE2022}. 

Models were trained with AMSGrad variant of Adam, with default parameters of $\beta_{1}$ = 0.9, $\beta_{2}$ =
0.999, and $\epsilon$ = $10^{-8}$. We used a learning rate of 0.01 and a batch size of 5. The learning rate was
reduced using an on-plateau scheduler based on the validation loss with a patience of 50 and a decay
factor of 0.8. 

\clearpage

\theoremstyle{remark}
\newtheorem{remark}{Remark}

\setcounter{secnumdepth}{4}
\setcounter{subsection}{0}
\renewcommand{\thesubsection}{\arabic{subsection}}
\renewcommand{\theparagraph}{\thesubsection.\arabic{paragraph}}
\makeatletter\renewcommand{\p@paragraph}{}\makeatother
\section*{ERRATUM}

\bigskip 

\noindent
The Supplement of the Letter claims that the parameter tensor of a linear
atomic cluster expansion (ACE) model is symmetric under joint permutation of its
channel and degree indices, and hence admits an exact \emph{symmetric} rank-one
(canonical-polyadic, CP) expansion at finite rank $K$ [\Leqs{8}{9} and the
Supplemental Material]. This is the basis of the ``tensor decomposition''
features $\tB_{i,k\vl\eta}$ of \Leq{12} and of the corresponding row of Table~I
of the Letter; we refer to these tensor-reduced ACE features as TRACE features below.
The claim is correct for correlation order $\nu \leq 3$, which covers all fitted models used for numerical experiments in the Letter, but it is
false for $\nu \geq 4$. The purpose of this erratum is to state precisely where
the flaw in the derivation is, and what the consequences are for the TRACE
model. We also briefly comment on equivariant predictions, which are not within
the scope of the original manuscript but of direct relevance due to the close
connection between the TRACE and MACE~\cite{MACE2022} models.

\subsection{Review of the incorrect result}

We follow the notation of the Letter and, as in its Supplemental Material,
absorb the element index $z$ into the channel index $n$. Equation numbers
prefixed by ``L'', such as \Leq{5}, refer to the Letter; unprefixed numbers
refer to the present erratum. The one-particle
basis $\phi_{nlm}$ is pooled over neighbours to give $A_{i,nlm}$, products over
$\nu$ slots give $\bm{A}_{i,\vn\vl\vm} = \prod_{t=1}^{\nu} A_{i,n_t l_t m_t}$,
and the invariant basis is
\begin{equation}
  B_{i,\vn\vl\eta}
  = \sum_{\vm} \CG^{\vl\eta}_{\vm}\, \bm{A}_{i,\vn\vl\vm},
  \qquad \eta = 1, \dots, d(\vl),
  \label{eq:B}
\end{equation}
where $\{\CG^{\vl\eta}\}_{\eta}$ is an orthonormal basis of the space
\begin{equation}
  \calK^{\vl}
  = \Big\{ \CG : \sum_{\vm} \CG_{\vm} \prod_{t} D^{l_t}_{m_t m_t'}(Q)
    = \CG_{\vm'} \ \ \forall Q \in \OO \Big\}
  \label{eq:Kl}
\end{equation}
of rotationally invariant couplings of the degrees $\vl = (l_1, \dots, l_\nu)$,
and $d(\vl) = \dim \calK^{\vl}$ is the number of independent couplings. A
complete linear model reads $\varphi_i = \sum_{\vn\vl\eta} c_{\vn\vl\eta}
B_{i,\vn\vl\eta}$, which is \Leq{5}; see \cite{ACE_ralf,DUSSON2022} for the
completeness. Because $\bm{A}_{i,\vn\vl\vm}$ is a product it is unchanged by any
joint permutation of the slots $(n_t, l_t, m_t) \mapsto (n_{\pi(t)}, l_{\pi(t)},
m_{\pi(t)})$, so in \Leq{5} only lexicographically ordered index tuples are
enumerated; the Supplemental Material instead retains all orderings and
works with a redundant basis.

The Letter then asserts (below \Leq{5}, and in the first section of the
Supplemental Material) that the coefficient tensor $c_{\vn\vl\eta}$ is
\emph{symmetric} under $(n_a, l_a) \leftrightarrow (n_b, l_b)$ at fixed $\eta$,
and concludes that it can be expanded in symmetric rank-one terms,
\begin{equation}
  c_{\vn\vl\eta}
  = \sum_{k=1}^{K} \lambda_{k\eta} \prod_{t=1}^{\nu} W^{k}_{n_t l_t},
  \label{eq:symcp}
\end{equation}
which is \Leq{9}, ``exactly for finite $K$, as $c$ is finite due to basis truncation''.
Substituting \eqref{eq:symcp} into \Leq{5} factorises the sum over $\vn$ and
yields the tensor-reduced features
\begin{equation}
  \tA_{i,klm} = \sum_{n} W^{kl}_{n} A_{i,nlm},
  \qquad
  \tB_{i,k\vl\eta} = \sum_{\vm} \CG^{\vl\eta}_{\vm}
    \prod_{t=1}^{\nu} \tA_{i,k l_t m_t},
  \label{eq:trace}
\end{equation}
which are \Leqs{10}{12}, in which the \emph{same} channel index $k$ appears in every
slot. The claim of exactness at finite $K$ is a claim that every linear ACE
model, \Leq{5}, can be written in the form \Leq{12}, that is, that the span of
the features \eqref{eq:trace} over all $K$ and all weights exhausts the span of
the ACE basis \eqref{eq:B}.

The Supplemental Material justifies the symmetry of $c_{\vn\vl\eta}$ by
observing that, once all orderings of the index tuple are retained, ``the
generalised Clebsch--Gordan coefficients do not depend on $\vn$''. That is true,
but it is not what is actually required. The symmetry claim tacitly assumes that
the coupling coefficients themselves are transported trivially along
permutations,
\begin{equation}
  \CG^{\pi\vl,\,\eta}_{\pi\vm} = \CG^{\vl\eta}_{\vm}
  \qquad \text{for all } \pi \in \Sym{\nu},
  \label{eq:assumed}
\end{equation}
with $(\pi\vm)_t = m_{\pi(t)}$ and likewise for $\vl$. For permutations that
change the ordering of $\vl$, i.e.\ $\pi\vl \neq \vl$, Eq.~\eqref{eq:assumed} is
harmless: it is simply a \emph{definition} of the couplings on the reordered
tuple. For permutations in the stabiliser
\begin{equation}
  \Stab(\vl) = \{ \pi \in \Sym{\nu} : l_{\pi(t)} = l_t \ \text{for all } t \},
\end{equation}
which is nontrivial whenever $\vl$ has repeated entries, Eq.~\eqref{eq:assumed}
is no longer a definition but a \emph{constraint} on the couplings: it demands
that every $\CG^{\vl\eta}$ be invariant under relabelling of the $\vm$ indices
by $\pi \in \Stab(\vl)$. This constraint holds for $\nu \leq 3$ (given
inversion symmetry; see paragraph~\ref{sec:nu4}) and fails in general for
$\nu \geq 4$.

\subsection{The correct relationship}
\label{sec:correct}

Let $\pi \in \Stab(\vl)$. Substituting $\vm \mapsto \pi\vm$ in the invariance
condition \eqref{eq:Kl} and relabelling the product over $t$, using
$l_{\pi(t)} = l_t$, shows that the relabelled coupling
$\vm \mapsto \CG^{\vl\eta}_{\pi\vm}$ again belongs to $\calK^{\vl}$. Since
$\{\CG^{\vl\eta}\}_\eta$ is a basis of that space,
\begin{equation}
  \CG^{\vl\eta}_{\pi\vm}
  = \sum_{\eta'} R^{\vl\pi}_{\eta\eta'}\, \CG^{\vl\eta'}_{\vm},
  \qquad \pi \in \Stab(\vl),
  \label{eq:recombination}
\end{equation}
for a uniquely determined matrix $R^{\vl\pi} \in \R^{d(\vl) \times d(\vl)}$.
Relabelling permutes the terms of the inner product
$\sum_{\vm} \CG_{\vm} \CG'_{\vm}$, so $R^{\vl\pi}$ is orthogonal, and
$\pi \mapsto R^{\vl\pi}$ is a (real, orthogonal) representation of the finite
group $\Stab(\vl)$ on the coupling space $\calK^{\vl}$. The assumption
\eqref{eq:assumed} is precisely the statement that this
representation is trivial, $R^{\vl\pi} = I$ for all $\pi \in \Stab(\vl)$.

Two consequences follow. First, for the basis functions \eqref{eq:B}, a change
of summation variable gives
\begin{equation}
  B_{i,\pi\vn,\vl,\eta}
  = \sum_{\eta'} R^{\vl\pi}_{\eta\eta'}\, B_{i,\vn\vl\eta'},
  \qquad \pi \in \Stab(\vl):
  \label{eq:B-perm}
\end{equation}
permuting the channel indices within slots of equal degree need \emph{not}
in general reproduce the same basis function; it \emph{recombines} the couplings. Second,
for the assembled invariant tensor
$T_{\vn\vl\vm} = \sum_\eta c_{\vn\vl\eta} \CG^{\vl\eta}_{\vm}$, which is what is
contracted against the symmetric product $\bm{A}_{i,\vn\vl\vm}$ and may
therefore be taken symmetric under joint permutation of $(\vn, \vl, \vm)$
without loss, the condition $T_{\pi\vn,\vl,\pi\vm} = T_{\vn\vl\vm}$ together
with \eqref{eq:recombination} and orthogonality of $R^{\vl\pi}$ yields
\begin{equation}
  c_{\pi\vn,\,\vl,\,:} = R^{\vl\pi}\, c_{\vn\vl:},
  \qquad \pi \in \Stab(\vl),
  \label{eq:c-correct}
\end{equation}
where $c_{\vn\vl:} \in \R^{d(\vl)}$ collects the coefficients over $\eta$.
This is the actual relationship between coefficients whose labels differ by a
permutation of the stabiliser. The coefficient tensor is not symmetric in
$(n_t, l_t)$ at fixed $\eta$; it is \emph{equivariant} under $\Stab(\vl)$, with
the coupling index transforming by $R^{\vl\pi}$. 

\paragraph{Consequence for the symmetric CP expansion.}
Any tensor of the form \eqref{eq:symcp} satisfies
$c_{\pi\vn,\vl,\eta} = c_{\vn\vl\eta}$ for $\pi \in \Stab(\vl)$: the factors are
indexed by the degree $l_t$ alone and cannot distinguish two slots carrying the
same degree. Combined with \eqref{eq:c-correct} this forces
\begin{equation}
  c_{\vn\vl:} \in \Fix(\vl)
  := \big\{ v \in \R^{d(\vl)} : R^{\vl\pi} v = v \ \ \forall \pi \in \Stab(\vl) \big\}
  \qquad \text{for every } \vn .
  \label{eq:fix}
\end{equation}
Therefore Eq.~\eqref{eq:symcp} can reproduce only those ACE models whose
coupling coefficients lie, for every $\vn$, in the subspace of couplings fixed
by the stabiliser. This restriction is independent of $K$ and of the choice of
weights $W$; increasing the rank does not help. The dimension of the fixed
subspace is given by the character formula
\begin{equation}
  \dim \Fix(\vl)
  = \frac{1}{|\Stab(\vl)|} \sum_{\pi \in \Stab(\vl)} \tr R^{\vl\pi},
\end{equation}
and the number of couplings lost per degree block is
$d(\vl) - \dim \Fix(\vl)$.

The same conclusion can be read off directly from the features
\eqref{eq:trace}. Because the same channel $k$ appears in every slot, the
product $\prod_t \tA_{i,k l_t m_t}$ is invariant under $\vm \mapsto \pi\vm$ for
$\pi \in \Stab(\vl)$, so in the contraction with $\CG^{\vl\eta}$ only the
stabiliser average
$P^{\vl}\CG := |\Stab(\vl)|^{-1} \sum_{\pi} \CG_{\pi\,\cdot}$ survives:
\begin{equation}
  \tB_{i,k\vl\eta}
  = \sum_{\vm} \big( P^{\vl} \CG^{\vl\eta} \big)_{\vm}
    \prod_{t} \tA_{i,k l_t m_t} .
  \label{eq:projected}
\end{equation}
If $\CG^{\vl\eta}$ lies in the orthogonal complement of $\Fix(\vl)$ then
$P^{\vl}\CG^{\vl\eta} = 0$ and $\tB_{i,k\vl\eta}$ vanishes identically, for
every channel $k$, every configuration, and every choice of weights.

A concrete case, $\vl = (1,1,2,2)$, is worked out in detail in
Section~\ref{sec:1122}. Section~\ref{sec:equivariant} briefly discusses the
equivariant case.

\begin{remark}
The Supplemental Material of the Letter already notes, in the section
``Uncoupled basis as a tensor sketch'', that when $\vl$ has repeated entries the
features of \Leq{12} are ``not a true tensor sketch'' because the expectation of
products of weights acquires additional terms. That observation has the same
root cause as the present erratum, namely identical weights in slots of equal
degree, but it was recorded there only as a bias in the inner-product
estimator, not as a loss of completeness.
\end{remark}

\paragraph{The problem first appears at $\nu = 4$.}
\label{sec:nu4}

For invariant, parity-even targets the assumption \eqref{eq:assumed} is in fact
correct for $\nu \leq 3$, so the argument of the Letter, and the exactness of
\Leq{9}, hold in that regime.

\begin{itemize}
\item $\nu = 2$. The only admissible degrees are $\vl = (l, l)$, with the single
  coupling $\CG_{m_1 m_2} = (-1)^{m_1}\delta_{m_1,-m_2}/\sqrt{2l+1}$, which is
  symmetric under the transposition $(1\,2)$, ensuring that $R^{\vl (12)} = 1$. 
\item $\nu = 3$. The triangle rule makes $d(\vl) \leq 1$ for every $\vl$, and
  the coupling is the Wigner $3j$ symbol, which under an odd permutation of its
  columns acquires the sign $(-1)^{l_1 + l_2 + l_3}$. Inversion symmetry of the
  target requires $l_1 + l_2 + l_3$ to be even, so the sign is $+1$ and the
  coupling is symmetric under the whole of $\Stab(\vl)$.
\end{itemize}
In both cases $R^{\vl\pi} = I$ on $\Stab(\vl)$, hence $\Fix(\vl) = \calK^{\vl}$,
and no basis functions are lost. Since the linear ACE fits reported in the Letter
(Figs.~2 and 4) use $\nu_{\max} = 3$ and the SOAP power spectrum corresponds to
$\nu = 2$, those experiments lie within the regime where the tensor-reduced
basis is complete. The restriction to parity-even targets is required here. 

The rank test in Fig.~1 of the Letter uses a standard ACE basis truncated by polynomial degree with $\nu \le 3$ and TRACE features with $\nu \le 5$. The plotted design-matrix ranks are measured quantities, and no comparison is made on the maximum rank attainable between TRACE and ACE. The qualitative conclusion, that the row rank of the design matrix grows near-linearly with the number of TRACE features up to full row rank, is unchanged. Had the complete ACE basis been included up to $\nu = 4$, a comparison of the maximum ranks could have revealed a discrepancy, since for $\nu \ge 4$ the TRACE features span a proper subspace of the ACE function space.

For $\nu \geq 4$ the multiplicity $d(\vl)$ exceeds one as soon as $\vl$ has
repeated entries that can be coupled through more than one intermediate, and
the stabiliser then acts nontrivially, as in Section~\ref{sec:1122}. Numerical tests show losses of $1$ of $3$ (e.g.\ $(1,1,2,2)$) and $2$ of $3$
($(1,1,1,1)$) and $4$ of $5$ ($(2,2,2,2)$) at $\nu = 4$, and up to $15$ of $16$
at $\nu = 5$ ($(2,2,2,2,2)$).

\subsection{Conclusions \& Outlook}

\begin{enumerate}
\item[(i)] The obvious remedy is the \emph{non-symmetric} tensor sketch that the
  Letter also introduces, \Leqs{14}{15} and the ``tensor sketch'' row of
  Table~I, in which each slot $t$ receives its own independent weight matrix
  $W^t$:
  \begin{equation}
    \hA_{it,klm} = \sum_{n} W^{tk}_{n} A_{i,nlm},
    \qquad
    \hB_{i,k\vl\eta} = \sum_{\vm} \CG^{\vl\eta}_{\vm}
      \prod_{t=1}^{\nu} \hA_{it,k l_t m_t} .
  \end{equation}
  The product is then no longer invariant under $\vm \mapsto \pi\vm$ for
  $\pi \in \Stab(\vl)$, so no coupling is annihilated in the contraction.
  Equivalently, the corresponding coefficient tensor
  $c_{\vn\vl\eta} = \sum_k \lambda_{k\vl\eta} \prod_t W^{tk}_{n_t}$ is an
  ordinary (non-symmetric) CP decomposition, which \emph{is} exact at finite
  $K$ for any finite tensor. The basis size remains $O(K)$; the cost is $\nu$
  sketches of the embedding instead of one. Any standard non-symmetric CP
  decomposition of the coefficient tensor, with learned rather than random
  factors, has the same property.
\item[(ii)] The tensor-decomposition features of \Leq{12} remain well defined
  and rotationally invariant; they parametrise a proper subspace of the ACE
  function space for $\nu \geq 4$ (paragraph~\ref{sec:nu4}), and of the equivariant ACE function space
  already for $\nu \geq 2$ (Section~\ref{sec:equivariant}). They are the product basis of the MACE
  architecture~\cite{MACE2022}, where the weights are learned and the construction is
  applied to the layer-to-layer maps of a message-passing network rather than
  to the final model. The practical consequences for the MACE architecture are
  unclear and are the subject of ongoing research.
\end{enumerate}

\subsection{The special case \texorpdfstring{$\vl = (1,1,2,2)$}{l = (1,1,2,2)}}
\label{sec:1122}

The simplest instance in which the symmetric expansion is incomplete for a
scalar, parity-even target -- the setting of an energy model under $\OO$ -- and
in which the stabiliser acts by pure signs, so that everything can be checked by
hand, is $\nu = 4$ with $\vl = (1,1,2,2)$.

\paragraph{The coupling space.}
The stabiliser is $\Stab(\vl) = \langle (1\,2), (3\,4) \rangle \cong
\mathbb{Z}_2 \times \mathbb{Z}_2$. Invariant couplings are enumerated by the
intermediate angular momentum $\lambda$ to which the two $l=1$ slots are coupled
and, separately, the two $l=2$ slots; reaching $L=0$ forces the two
intermediates to agree. There are three couplings,
\begin{equation}
  \CG^{(\lambda)}_{m_1 m_2 m_3 m_4}
  = \sum_{\mu\mu'}
    C^{\lambda\mu}_{1 m_1,\, 1 m_2}\;
    C^{\lambda\mu'}_{2 m_3,\, 2 m_4}\;
    C^{00}_{\lambda\mu,\, \lambda\mu'},
  \qquad \lambda \in \{0, 1, 2\},
  \label{eq:1122-couplings}
\end{equation}
with $C^{\lambda\mu}_{l_1 m_1, l_2 m_2}$ the ordinary Clebsch--Gordan
coefficients, so $d(\vl) = 3$. The exchange symmetry of the Clebsch--Gordan coefficients,
$C^{\lambda\mu}_{l m_1,\, l m_2} = (-1)^{2l - \lambda} C^{\lambda\mu}_{l m_2,\, l m_1}
= (-1)^{\lambda} C^{\lambda\mu}_{l m_2,\, l m_1}$, applied to
\eqref{eq:1122-couplings} gives
\begin{equation}
  \CG^{(\lambda)}_{(1\,2)\vm} = (-1)^{\lambda}\, \CG^{(\lambda)}_{\vm},
  \qquad
  \CG^{(\lambda)}_{(3\,4)\vm} = (-1)^{\lambda}\, \CG^{(\lambda)}_{\vm}.
\end{equation}
In the basis ordered by $\lambda = 0, 1, 2$,
\begin{equation}
  R^{\vl,(1\,2)} = R^{\vl,(3\,4)} = \operatorname{diag}(1, -1, 1),
  \qquad
  R^{\vl,(1\,2)(3\,4)} = I .
\end{equation}
In particular, the coupling $\CG^{(1)}$ is antisymmetric under each
transposition. (We have also confirmed these calculations numerically by
constructing $\calK^{\vl}$ and the matrices $R^{\vl\pi}$ explicitly.)

\paragraph{The antisymmetric coupling cannot be reached.}
Since the matrices $R^{\vl\pi}$ are diagonal, the fixed subspace in
\eqref{eq:fix} is $\Fix(\vl) = \spn\{\CG^{(0)}, \CG^{(2)}\}$, and $\CG^{(1)}$
spans its orthogonal complement. The stabiliser average in
\eqref{eq:projected} is
$P^{\vl}\CG^{(1)} = \tfrac14 (1 - 1 - 1 + 1)\, \CG^{(1)} = 0$, hence the
corresponding tensor-reduced feature vanishes. This can also be seen directly:
writing $\tB^{(1)}_{i,k}$ for the feature \eqref{eq:trace} with coupling
$\CG^{(1)}$, and relabelling $m_1 \leftrightarrow m_2$ in the sum, we obtain
\begin{align}
  \tB^{(1)}_{i,k}
  &= \sum_{\vm} \CG^{(1)}_{m_1 m_2 m_3 m_4}\,
    \tA_{i,k1m_1} \tA_{i,k1m_2} \tA_{i,k2m_3} \tA_{i,k2m_4} \notag \\
  &= \sum_{\vm} \CG^{(1)}_{m_2 m_1 m_3 m_4}\,
    \tA_{i,k1m_1} \tA_{i,k1m_2} \tA_{i,k2m_3} \tA_{i,k2m_4}
  = - \tB^{(1)}_{i,k},
\end{align}
that is, $\tB^{(1)}_{i,k} = 0$ for every $k$, every configuration and every
choice of the weights $W$: an antisymmetric coupling is contracted against a
product that is symmetric in $m_1 \leftrightarrow m_2$.

The same conclusion is obtained at the level of the ACE parameters. Write
$B^{(1)}_{i,\vn}$ and $c^{(1)}_{\vn}$ for the basis functions \eqref{eq:B} and
the coefficients belonging to $\CG^{(1)}$. By \eqref{eq:B-perm}, where the entry
of $R^{\vl,(1\,2)}$ belonging to $\lambda = 1$ is $-1$, the basis functions are antisymmetric in the two
$l = 1$ channels,
$B^{(1)}_{i,n_2 n_1 n_3 n_4} = - B^{(1)}_{i,n_1 n_2 n_3 n_4}$; they vanish for
$n_1 = n_2$ but are in general nonzero otherwise. Collecting the terms
$(n_1, n_2)$ and $(n_2, n_1)$, the contribution of this coupling to an ACE model
is
\begin{equation}
  \sum_{\vn} c^{(1)}_{\vn} B^{(1)}_{i,\vn}
  = \sum_{n_1 < n_2} \sum_{n_3, n_4}
    \big( c^{(1)}_{n_1 n_2 n_3 n_4} - c^{(1)}_{n_2 n_1 n_3 n_4} \big)\,
    B^{(1)}_{i,n_1 n_2 n_3 n_4},
  \label{eq:1122-antisym}
\end{equation}
so only the part of $c^{(1)}$ that is antisymmetric in $(n_1, n_2)$ enters the
model, in agreement with \eqref{eq:c-correct}. For the symmetric expansion
\eqref{eq:symcp},
\begin{equation}
  c^{(1)}_{n_1 n_2 n_3 n_4} - c^{(1)}_{n_2 n_1 n_3 n_4}
  = \sum_{k=1}^{K} \lambda_{k}\,
    \big( W^{k}_{n_1 1} W^{k}_{n_2 1} - W^{k}_{n_2 1} W^{k}_{n_1 1} \big)\,
    W^{k}_{n_3 2} W^{k}_{n_4 2}
  = 0
\end{equation}
for every $K$, $\lambda$ and $W$, and the right-hand side of
\eqref{eq:1122-antisym} vanishes. Conversely, any ACE model with
$c^{(1)}_{n_1 n_2 n_3 n_4} = - c^{(1)}_{n_2 n_1 n_3 n_4} \neq 0$, which is the
generic case, has no representation of the form \eqref{eq:symcp} or
\eqref{eq:trace}; the expansion \Leq{9} is therefore not exact for any
finite $K$.

\subsection{Equivariant targets (MACE)}
\label{sec:equivariant}

The Letter concerns invariant models, but the same product basis can be used
with $L \geq 1$ couplings to build equivariant features, for instance the
equivariant messages of MACE~\cite{MACE2022}. There the protection afforded by $\nu \leq 3$
disappears, because the multiplicity-one argument of
paragraph~\ref{sec:nu4} is a statement about the \emph{trivial} representation
only. We record the two smallest instances without proof; both follow from
\eqref{eq:fix} exactly as above.

\begin{itemize}
\item \emph{Vectors, $\nu = 3$.} For a polar vector target ($L = 1$, odd
  parity, e.g.\ a dipole) and $\vl = (1,1,1)$ there are three couplings,  permuted
  by $\Stab(\vl) = \Sym{3}$. Only their symmetric sum is fixed; with a single
  sketched channel all three collapse to $|\tA_{i,k}|^2 \tA_{i,k}$, so one of
  three couplings is reachable via TRACE. 
\item \emph{Pseudovectors, $\nu = 2$.} For an axial vector target ($L = 1$,
  even parity, e.g.\ a magnetic moment) and $\vl = (1,1)$ the single coupling is
  the cross product $a \times b$, which is antisymmetric. With a single sketched
  channel it becomes $\tA_{i,k} \times \tA_{i,k} = 0$: the entire degree block
  is absent from the tensor-reduced basis. 
\end{itemize}
Thus for equivariant outputs the symmetry clash occurs at $\nu = 3$ for vectors
and already at $\nu = 2$ for pseudovectors.

\end{document}

%% file: acronyms.tex
\DeclareAcronym{DFT}{
short = DFT,
long = Density-Functional Theory
}

\DeclareAcronym{SOAP}{
short = SOAP,
long = Smooth Overlap of Atomic Positions
}

\DeclareAcronym{ACE}{
short = ACE,
long = Atomic Cluster Expansion
}

\DeclareAcronym{MACE}{
short = MACE,
long = Multi-ACE
}

\DeclareAcronym{RR}{
short = RR,
long =  Ridge Regression
}

\DeclareAcronym{KRR}{
short = KRR,
long = Kernel Ridge Regression
}

\DeclareAcronym{GAP}{
short = GAP,
long = Gaussian Approximation Potential
}

\DeclareAcronym{II}{
short = II,
long = Information Imbalance
}

\DeclareAcronym{RP}{
short = RP,
long = Random Projection
}

\DeclareAcronym{JL}{
short = JL,
long = Johnson-Lindenstrauss
}

\DeclareAcronym{TR}{
short = TR,
long = tensor-reduced
}

\DeclareAcronym{MPNN}{
short = MPNN,
long = Message Passing Neural Network
}

\DeclareAcronym{CLSR}{
short = CLSR,
long = Compressed Least-Square Regression
}